\RequirePackage{fix-cm}
\documentclass[smallextended]{svjour3}

\smartqed  
\usepackage{cite}
\usepackage{amsmath,amssymb,amsfonts}
\usepackage{algorithmic}
\usepackage{graphicx}
\usepackage{textcomp}
\usepackage{xcolor}
\usepackage{multirow}

\PassOptionsToPackage{hyphens}{url}
\usepackage{hyperref}

\usepackage{listings}
\usepackage{comment}
\usepackage{enumitem}
\usepackage{booktabs}
\usepackage{soul}
\usepackage{xcolor}

\newcommand{\change}[1]{\textcolor{black}{#1}}

\usepackage{longtable}
\usepackage{fancybox}
\usepackage{tikz,lipsum,lmodern}
\usepackage[most]{tcolorbox}

\begin{document}

\title{Software Development with Feature Toggles: Practices used by Practitioners
}
%\subtitle{Do you have a subtitle?\\ If so, write it here}

%\titlerunning{Short form of title}        % if too long for running head

\author{Rezvan Mahdavi-Hezaveh         \and
        Jacob Dremann                  \and
        Laurie Williams
}

%\authorrunning{Short form of author list} % if too long for running head

\institute{Rezvan Mahdavi-Hezaveh \at
              Department of Computer Science, North  Carolina State University, Raleigh, NC, USA\\
              \email{rmahdav@ncsu.edu} 
           \and
           Jacob Dremann \at
              Department of Computer Science, North  Carolina State University, Raleigh, NC, USA\\
              \email{jtdreman@ncsu.edu}
           \and
           Laurie Williams \at
              Department of Computer Science, North  Carolina State University, Raleigh, NC, USA\\
              \email{laurie\_williams@ncsu.edu}
}

\date{Received: date / Accepted: date}
% The correct dates will be entered by the editor

\maketitle

\begin{abstract}
Background: Using feature toggles is a technique that allows developers to either turn a feature on or off with a variable in a conditional statement. Feature toggles are increasingly used by software companies to facilitate continuous integration and continuous delivery. 
%Using feature toggles allows developers to integrate incomplete features into the code base without impacting user behavior and to run experiments on new features in an application. 
However, using feature toggles inappropriately may cause problems which can have a severe impact, such as code complexity, dead code, and system failure. For example, the erroneous repurposing of an old feature toggle caused Knight Capital Group, an American global financial services firm, to go bankrupt due to the implications of the resultant incorrect system behavior.  \\ 
Aim: The goal of this research project is to aid software practitioners in the use of practices to support software development with feature toggles through an empirical study of feature toggle practice usage by practitioners. \\
Method: We conducted a qualitative analysis of 99 artifacts from the grey literature and 10 peer-reviewed papers about feature toggles.  We conducted a survey of practitioners from 38 companies. \\
Results: We identified 17 practices in 4 categories: Management practices, Initialization practices, Implementation practices, and Clean-up practices. We observed that all of the survey respondents use a dedicated tool to create and manage feature toggles in their code. Documenting feature toggle's metadata, setting up the default value for feature toggles, and logging the changes made on feature toggles are also frequently-observed practices.\\
 Conclusions: The feature toggle development practices discovered and enumerated in this work can help practitioners more effectively use feature toggles. This work can enable future mining of code repositories to automatically identify feature toggle practices.
%researches about feature toggles since the result of qualitative analysis researches could be the start point to conduct following quantitative analysis researches in the same domain. Also some of these identified feature toggle practices could be identified automatically by analysis the code repository of projects.  
\end{abstract}

\section{Introduction}
\label{intro}
In 2012, developers in Knight Capital Group, an American global financial services firm, updated their automated, high-speed, algorithmic router which inadvertently repurposed a feature toggle\footnote{Feature toggles are also called feature flags, feature bits, feature flippers and feature switches \cite{fowlerfeaturetoggle}.}, activating functionality which had been unused for 8 years. Within 2 minutes, developers realized the deployed code behaved incorrectly but took 45 minutes to stop the system. During that time, Knight Capital lost nearly 400 million dollars, which caused the group to go bankrupt \cite{knightcapital}. As illustrated, using feature toggles without following good practices can be detrimental to an organization.

Developers guard blocks of code with a variable as a feature toggle in conditional statements, and by changing the value of the variable, enable or disable that part of the code in the system's execution. The value of the variable could be changed either in the code or remotely on a configuration server. \change{Feature toggles have similarities to configuration options. However, as will be discussed in Section \ref{background-relatedworks}, feature toggles also have significant differences to configuration options \cite{meinicke2019exploring}}. The use of feature toggles is a technique often used in continuous integration (CI) and continuous delivery (CD) contexts that allows teams to incrementally integrate and test a new feature even when the feature is not ready to be released \cite{rahman2016feature}\cite{parnin2017top}. Developers also use feature toggles for other purposes, such as gradual roll out and experiments. However, feature toggles can turn into technical debt~\cite{birdtechnicaldebt}. Using feature toggles adds more decision points to the code which adds more complexity. This increased complexity drives the need to remove toggles when their purpose is complete. 

 \change{A software development \textit{practice} is an activity or step carried out to achieve a goal during the development of software. For example, unit testing is a practice for white-box testing of implementation code.} The identification and categorization of feature toggle practices used in industry may help software practitioners to use toggles more efficiently and to control the accumulation of technical debt.  \textit{The goal of this research project is to aid software practitioners in the use of practices to support software development with feature toggles through an empirical study of feature toggle practice usage by practitioners.}  Software practitioners prefer to learn through the experiences of other software practitioners \cite{moore2009crossing}.  As such, our study obtains practice usage from practitioners.

We state the following research questions:
\begin{description}
\item[RQ1]{(Identification): What are the feature toggle practices that software practitioners use?}
\item[RQ2]{(Frequency): How frequently are feature toggle practices used?}
\end{description}

To answer the first research question, using a keyword search we collected peer-reviewed papers and artifacts from the grey literature about feature toggles. We used an open coding technique \cite{saldana2015coding} to perform qualitative analysis on these artifacts to identify practices. \change{We do not call the identified practices ``best'' practices because we did not obtain enough evidence to select any practice as a ``best'' practice}. To answer the second research question, we analyzed company-specific peer-reviewed papers and artifacts from the grey literature and conducted a survey to find the frequency of usage of the identified practices. 

%including 10 peer-reviewed papers, 41 blog posts and online articles, and 15 videos

We summarize the contribution of this paper:
\begin{enumerate}
\item A list of 17 practices in four categories used to support software development with feature toggle; and
\item An analysis of the the frequency of usage of feature toggle practices in industry.
\end{enumerate}

The rest of the paper is organized as follows: in Section \ref{background-background}, we describe the background of the research area. In Section \ref{background-relatedworks}, we briefly describe prior academic work related to our paper. In Section \ref{methodology}, we explain our research methodology. In Section \ref{result}, we report our findings. In Section \ref{Discussion}, we discuss our findings. We enumerate the limitations of our study in Section \ref{limitation}. We conclude and describe future work on feature toggles in Section \ref{conclusion}.

\section{Background} \label{background-background}
In this section, we first provide briefly the definitions of CI and CD. Next, we explain the feature toggle concept and it's types. At the end, we describe the grey literature and its importance for our paper.

\subsection{Continuous Integration(CI) and Continuous Delivery(CD)} \label{CICD}

Companies must deliver valuable software rapidly to be competitive. This expectation leads companies to use CI and CD to make development cycles shorter. CI is a practice of integrating and automatically building and testing software changes to the source repository after each commit \cite{humble2010continuous}. CD is a practice for keeping the software in a state such that it can be released to a production environment at any time \cite{fowlercontinuousdelivery}.  CI/CD refers to a combination of these two practices and enables delivering code changes frequently. Using feature toggles is one of the techniques that is used by numerous software companies who practice CI/CD \cite{parnin2017top}. 

\subsection{Feature Toggles} \label{featuretogglesection}
Programming languages have long provided the language constructs to implement feature toggles.  However, the first use of this language construct to support CI/CD was at Flickr in 2009 \cite{rossflippingout}. Figure \ref{fig:toggle} is an example of a feature toggle. In this example, the dynamic choice of a search algorithm depends on the value of the \textsf{useNewAlgorithm} toggle. If the value of this toggle is true, then the new search algorithm is used, otherwise the \textsf{Search} method calls the old search algorithm.

\begin{figure}
    \centering
    \begin{lstlisting}[xleftmargin=17pt,xrightmargin=17pt,frame=single]
function Search(){
    var useNewAlgorithm = false;
    if(useNewAlgorithm){
      return newSearchAlgorithm();
    }else{
      return oldSearchAlgorithm();
    }
}
    \end{lstlisting}
    \caption{An example of a feature toggle.}
    \label{fig:toggle}
\end{figure}

Feature toggles have been categorized into five types in software systems in \cite{rahman2016feature} and  \cite{hodgsonfeaturetoggle}: 
\begin{itemize}
    \item \textit{Release toggles}: Toggles used to add new features in a trunk-based development context. In trunk-based development, all developers commit changes to one shared branch. Using release toggles in trunk-based development supports CI/CD for partially-completed features \cite{hodgsonfeaturetoggle} \cite{rahman2016feature}.
    \item \textit{Experiment toggles}: Toggles used to perform experimentation on the software, such as is done by Microsoft \cite{microsoftexperiment} \cite{kohavi2009controlled}, to evaluate new features changes and their influence on user-observable behavior \cite{hodgsonfeaturetoggle}.
    \item \textit{Ops toggles}: Toggles used to control the operational aspect of the system behavior. When a new feature is deployed, system operators can disable the feature quickly if it performs unexpectedly \cite{hodgsonfeaturetoggle}.
    \item \textit{Permission toggles}: Toggles used to provide the appropriate functionality to a user, e.g. special features for premium or paid users \cite{hodgsonfeaturetoggle}. Permission toggles also called long-term business toggles in \cite{rahman2016feature}.
    \item \textit{Development Toggles}: Toggles used for enabling or disabling certain features to test and debug code \cite{rahman2016feature}.
\end{itemize}

Permission toggles, ops toggles and development toggles are long-lived toggles based on their usage purpose in the code.  Release toggles and experiment toggles are short-lived toggles \cite{hodgsonfeaturetoggle} \cite{rahman2016feature}.

\change{\subsection{Grey Literature}}
\change{Grey literature is defined as \textit{``... literature that is not formally published in sources, such as books or journal articles''} \cite{lefebvre2008searching}. Software practitioners may share their experiences as grey literature which can be considered as a valuable resource for researchers \cite{garousi2019benefitting} and other practitioners. Academic publications reflect the state-of-the-art and grey literature provide insight to the state-of-the-practice in any research area. In practical research areas such as software engineering, combining the state-of-the-art and state-of-the-practice is important to provide valuable results  \cite{garousi2016need}. In the area of feature toggles, a large number of grey literature artifacts exist but only a small number of peer-reviewed papers have been published.  As we will discuss in Step Two of the methodology in Section~\ref{methodology}, we use the quality assessment checklist of grey literature for software engineering provided in \cite{garousi2019guidelines} to evaluate the quality of our grey literature artifacts.}

%\change{As the usage of feature toggles is increasing over time}, tools have been developed to help developers use toggles more efficiently. For example, the LaunchDarkly\footnote{\url{https://launchdarkly.com/}} feature management platform helps practitioners to create new feature toggles, change their status, track their changes, and control their life cycle. Feature toggle libraries in programming languages, such as Java, JavaScript, Ruby, Python, and PHP, can also be used to manage feature toggles. These libraries can be added to the code to aid in the creation, management, and use of feature toggles.

\section{Related Work} \label{background-relatedworks}

Rahman et al. \cite{rahman2015synthesizing} performed a qualitative grey literature study and conducted follow-up inquiries to study continuous deployment practices. They reported 11 continuous deployment practices used by 19 software companies. Using feature toggles is one of these 11 practices that is used by 13 of the 19 companies. In addition, at the Continuous Deployment Summit \cite{parnin2017top} 2015, researchers and practitioners from 10 companies shared their best practices and challenges. Parnin et al. \cite{parnin2017top} disseminated 10 best practices from the Summit, including the use of feature toggles to implement Dark Launches\footnote{ \textit{Dark launching} is a practice in which code is incrementally deployed into production but remains invisible to users \cite{parnin2017top}.}.

To understand the drawbacks, strengths, and cost of using feature toggles in practice, Rahman et al. \cite{rahman2016feature} performed a thematic analysis of videos and blog posts created by release engineers. %They reported the purpose of using feature toggles: rapid release,  trunk-based development, and A/B testing. 
They provided six themes founded in analyzed videos and blog posts, such as technical debt and combinatorial feature testing. To identify feature toggle practices, we used videos and blog posts from
\cite{rahman2016feature} and additional peer-reviewed papers and grey literature artifacts including more videos and blog posts we found. Rahman et al. \cite{rahman2016feature} also performed a quantitative analysis of feature toggle usage across 39 releases of Google Chrome from 2010 to 2015 and mined a spreadsheet used by Google developers for feature toggle maintenance. %They quantified the prevalence of three major types of feature toggles used in Chrome: development toggles (33\%), long-term business toggles (33\%), and release toggles (34\%). We mentioned four suggested types for feature toggles \cite{hodgsonfeaturetoggle} in Section \ref{background-background}. \change{We show the mapping between the suggested types of feature toggles in \cite{hodgsonfeaturetoggle} and identified types of feature toggles by Rahman et al. in Table \ref{mapping}.} %Among three feature toggle types in Chrome, release toggles are mapped to release toggles in suggested types, long-term business toggles are mapped to permission toggles and development toggles do not map to any suggested categories directly. Development toggles are used for testing and debugging but none of the suggested types point to this usage. 
Release toggles should be short-lived toggles but Rahman et al. observed that 53\% of the release toggles exist for more than 10 releases in Chrome. They classified unused but existing release toggles in the code as technical debt. \change{The goal of our study is to identify feature toggle practices and their usage in industry while the goal of their paper was to understand the drawbacks, strengths, and cost of using feature toggles.}

\iffalse
\change{
\begin{table}
    \centering
    \caption{The mapping between feature toggle types suggested in \cite{hodgsonfeaturetoggle} and \cite{rahman2016feature}}
    \label{mapping}    
    \begin{tabular}{l|l}
    \toprule
         Feature toggle type in \cite{hodgsonfeaturetoggle} & Feature toggle type in \cite{rahman2016feature} \\\midrule
         Release toggles & Release toggles \\
         Experiment toggles & - \\
         Ops toggles & - \\
         permission toggles & long-term business toggles \\
         - & development toggles \\
         \bottomrule
    \end{tabular}
\end{table}
}
\fi

Rahman et al. \cite{rahman2018modular} extracted four architectural representations of Google Chrome: 1) conceptual architecture; 2) concrete architecture; 3) browser reference architecture; and 4) feature toggle architecture. Using the extracted feature toggle architecture, developers can find out which feature effects which module and which module is affected by which feature. \change{The goal of their study was to show how developers can get a new viewpoint into the feature architecture of the system using the extracted feature toggle architecture.} Their result raise awareness of the impact of using feature toggles on the modular architecture of the system. \change{In our paper, we focus on the practices of using feature toggles which is not in the scope of their study.}

\change{Meinicke et al.  \cite{meinicke2019exploring} explored the differences and commonalities between configuration options and feature toggles by conducting nine semi-structured interviews with feature toggle experts. Configuration options are key-value pairs used by end users to include or exclude functionality in a software system. During the interviews, the authors asked practitioners about the existing literature on feature toggles. Then, they discussed with the interviewees the configuration options topics and asked them if they saw common challenges and solutions. % The authors identified the differences in 10 themes: Goals, Who makes configuration decisions?, Complexity, Removing configuration decisions, Feature traceability, Documentation, Constraints, Dependencies, Feature Interactions, and Testing. 
The researcher found that although feature toggles and configuration options are similar concepts, they have distinguishing characteristics and requirements. The goal of the usage and challenges of each of the techniques are distinct. The researcher identified 10 themes for the differences, such as their users (The feature toggles are used by developers but configuration options are used by end users) and their lifetime (The feature toggles will be removed from code ideally but configuration options can exist permanently).} %Meinicke et al.  claimed that there is a decade of research on configuration options in academia and a lot of information about feature toggles in industry, but both communities are unaware of each other. They suggested to transfer information from one community to the other.

\change{Sayagh et al. \cite{sayagh2018software} aimed to understand the process required by practitioners to aggregate configuration options in the software system, the challenges they face, and best practices that they could follow. To achieve their goal, the authors did 14 interviews with software engineering experts, conducted a survey on Java software engineers, and did a literature review on the academic papers in the area of configuration options. They identified 9 configuration management activities, 21 configuration challenges, and 24 expert recommendations. One of the reported challenges, the increasing complexity of the code by adding configuration options, is the same as the challenges of using feature toggles. In addition, eight of the 17 practices we identified, as will be discussed in Section \ref{practices}, have partial overlap with their reported activities, challenges recommendations, such as using naming conventions. However, because of the differences identified between feature toggles and configuration options in \cite{meinicke2019exploring}, ten practices out of 17 identified practices are not mentioned in \cite{sayagh2018software}. As an example, configuration options are intended to be ``permanent'' in the code but feature toggles intended to be ``temporary''.
This difference could explain the absence of our identified practices in the clean-up category in Sayegh et al. \cite{sayagh2018software}'s paper. As we provide each practice in Section~\ref{practices}, we will provide information on whether the feature toggle practice is in common with configuration options.}

None of this related work on feature toggles focus on identifying the feature toggle practices \change{broadly} used in industry and their usage frequency. We fill this gap in this paper.

\begin{figure*}[t]
\centerline{\includegraphics[width=360pt]{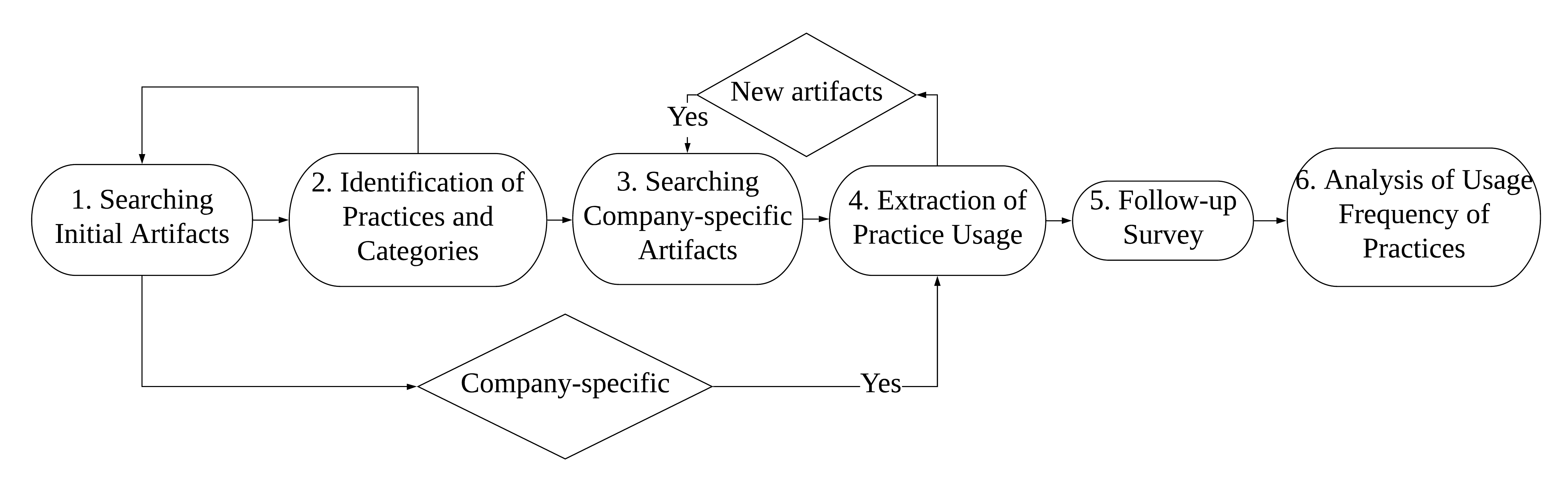}}
\caption{The research methodology.}
\label{fig:methodology}
\end{figure*}

\section{Research Methodology} \label{methodology}

We describe the steps of our methodology to answer the research questions. Our methodology has six steps, as shown in Figure~\ref{fig:methodology}. \change{We started by searching for initial set of peer-reviewed papers and artifacts from the grey literature related to our study scope, in Step One. In Step Two, we identified and categorized practices found in this literature, analyze the quality of grey literature artifacts and calculate the level of confidence for identified practices. In Steps Three and Four, we iteratively searched for grey literature related to specific companies and extract the usage of identified practices. Then, we sent a survey to practitioners, in Step Five. Finally in Step Six, we analyzed the usage frequency of practices and compare practices, if possible.} Each of these steps \change{from Figure~\ref{fig:methodology}} will be explained in detail in the following sub-sections.

\subsection{Step One: Searching Initial Artifacts}

The first step in our research methodology in Figure~\ref{fig:methodology} is to use a keyword search in the Google search engine to identify grey literature and in Google Scholar to find peer-reviewed papers. We used the following search terms: `feature toggle'; `feature flag'; `feature switch'; `feature flipper'; and `feature bit'. These search terms were obtained from Fowler's blog post \cite{fowlerfeaturetoggle}. 

\change{Selecting the related peer-reviewed papers and grey literature artifacts is done by first and second author of the paper. For grey literature, reviewing all the results of each searches in Google search engine was infeasible. The most relevant links are provided earlier in the Google search engine. We reviewed the first 10 pages of the results of each search to select the most relevant links. To determine if a link is related to the scope of our research, we read the article by looking for the search term and read 2--3 sentences before and after the search term in the text. For videos, we kept all the links and analyzed them in Step Two of the methodology.} In the relevant grey literature artifacts, we used a snowballing approach \cite{wohlin2014guidelines}. We clicked on links and the references to other feature toggle resources found in the artifact, and we read the papers, articles or watched the videos. \change{For peer-reviewed papers, we checked the titles of search result and kept those which were related to our topic. We used the snowballing approach for peer-reviewed papers. We checked the references and selected the related ones. \textit{In the rest of the paper, we use the term ``artifacts'' to refer to the set of ``peer-reviewed papers and grey literature artifacts''.}} \change{The time of the publishing of all collected artifacts is before June 2019.} Some collected artifacts in Step One were company-specific artifacts that were often written by a release manager or developer, referencing feature toggle usage at a specified company. We used these company-specific artifacts in Step Two and Step Four.

\begin{figure*}[t]
\includegraphics[width=\textwidth]{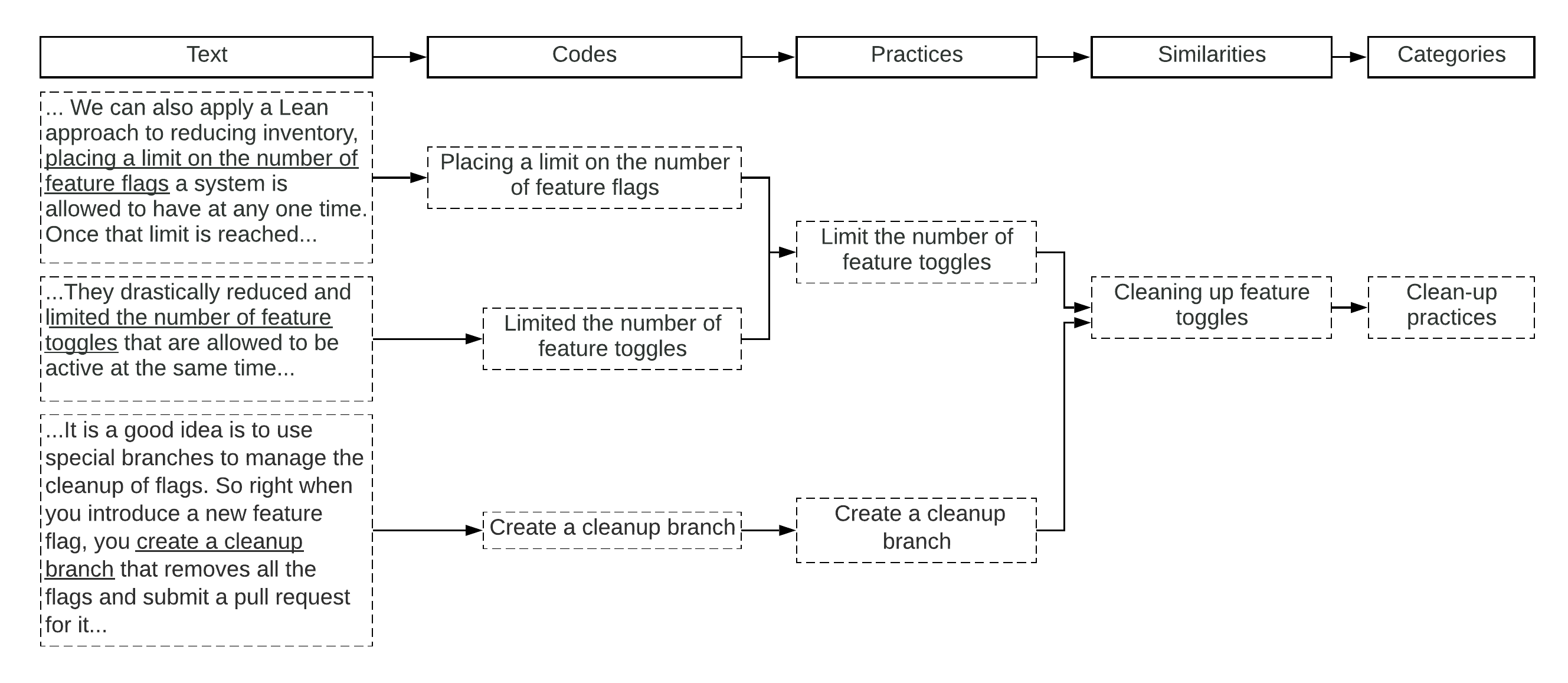}
\caption{An example of using the open coding technique.}
\label{fig:opencoding}
\end{figure*}

\subsection{Step Two: Identification of Practices and Categories}

The grey literature and peer-reviewed papers found in Step One were used to identify feature toggle practices. We analyzed the artifacts using an open coding technique, a technique to analyze textual data by coding (i.e. labeling) concepts and identifying categories based on similarity and dissimilarity of codes \cite{saldana2015coding}. First, we took notes from the videos. Then, we coded the suggested recommendations, experiences, and implementation details about using feature toggles mentioned by practitioners in the textual artifacts and in the notes of videos. \change{The coding of the artifacts was done by the first and second authors of the paper.}

After the identification of practices, we observed similarities and dissimilarities between practices. We put practices with similarities into one category based on an open coding technique and found four categories. We give an example of using open coding with a sample of our data in Figure~\ref{fig:opencoding}. In this figure, three paragraphs from three artifacts are shown and codes are assigned to them. The codes of the two first paragraphs pointed to the same concept so we grouped them as ``Limit the number of feature toggles''. The last code is changed to ``Create a cleanup branch'' practice. The similarity between these two extracted practices is pointing to cleaning up feature toggles, so the two practices are grouped as ``Clean-up practices''. The result of this step is the answer to the first research question (RQ1-Identification). 
\change{Step One and Step Two of the methodology were done iteratively. The snowball approach was terminated when we did not identify any new practices and did not find any new artifact.}

\change{After identification of practices from artifacts, we specified a ``Level of confidence'' for each practice which was used to quantify our confidence in the quality, correctness and importance of the identified practices. We used the quality assessment checklist of grey literature for software engineering provided in \cite{garousi2019guidelines} and shown in Table~\ref{checklist}. Following their example, each of the 20 question is assigned a score of 1, 0.5 or 0, so the highest score would be 20. We assigned a score of 20 to peer-reviewed papers.}

% \begin{table}[]
{\color{black}
\begin{longtable}{p{2cm}|p{4.2cm}|p{4.2cm}}
\caption{The quality assessment checklist of grey literature for software engineering adapted from \cite{garousi2019guidelines}.}
    \label{checklist} 
    \\
% \begin{tabular}{p{2cm}|p{4.2cm}|p{4.2cm}}
\hline
Criteria & Questions & Notes                                                \\ \hline

\multirow{4}{*}
{\begin{tabular}[c]{@{}l@{}} 
Authority of \\the producer
\end{tabular}}
& Is the publishing organization reputable? 
& The authorship is attributed to an organization.        
\\ \cline{2-3} 

& 
% \begin{tabular}[c]{@{}l@{}} 
Is an individual author associated with a reputable %\\ organization? 
% \end{tabular} 
& The authorship is attributed to an individual author(s).  
\\ \cline{2-3} 

& Has the author published other work in the field?  & 
Having other grey literature in the area.    
\\ \cline{2-3} 

& Does the author have expertise in the area?  
&  Having experience in the area.                     
\\ \hline

\multirow{6}{*}{Methodology} 
& Does the source have a clearly stated aim? 
& Having clear related subject.
\\ \cline{2-3} 

& Does the source have a stated methodology?  
& Having an structured flow for discussion.                      
\\ \cline{2-3} 

& 
% \begin{tabular}[c]{@{}l@{}} 
Is the source supported by authoritative, %\\ 
contemporary references? 
% \end{tabular} 
& Having any references. \\ \cline{2-3}

& Are any limits clearly stated? 
& Pointing to at least one related limitation.                  
\\ \cline{2-3} 

& Does the work cover a specific question? 
& Covering the concept of feature toggles.       
\\ \cline{2-3} 

& Does the work refer to a particular population or case?  
& Related to feature toggles.   
\\ \hline

\multirow{4}{*}{Objectivity} 
& Does the work seem to be balanced in presentation?  
& Discussing the subject from different views.               
\\ \cline{2-3} 

& Is the statement in the sources as objective as possible? & Including enough evidence.
\\ \cline{2-3} 

& Is there vested interest? 
&  Unbiased to any organization's tool.
\\ \cline{2-3} 

& Are the conclusions supported by the data?  
& Having reasonable conclusion.                      
\\ \hline

Date & Does the item have a clearly stated date?  
& Including date.                    
\\ \hline

% \begin{tabular}[c]{@{}l@{}} 
Position w.r.t. %\\ 
related sources 
% \end{tabular} 
& 
% \begin{tabular}[c]{@{}l@{}} 
Have key related grey literature or formal sources been %\\ 
linked to/discussed? 
% \end{tabular}  

& Having at least one of the key grey artifacts of the area. 
\\ \hline

\multirow{2}{*}{Novelty} 
& Does it enrich or add something unique to the research?  
& Including any valuable data for the research.                         \\ \cline{2-3} 

& Does it strengthen or refute a current position?  
& Supporting any valuable data for the research.      
\\ \hline

Impact &
%  \begin{tabular}[c]{@{}l@{}} 
Normalize Number of backlinks, Number of social media shares into a single metric 
% \end{tabular}  

& 
% \begin{tabular}[c]{@{}l@{}} 
For backlinks: %\\
\url{https://www.seoreviewtools.com/valuable-backlinks-checker/} 
%\\ 
, For social media share: %\\ 
\url{http://www.sharedcount.com/}
% \end{tabular}                                 
\\ \hline

Outlet type & % \begin{tabular}[c]{@{}l@{}}
\begin{itemize}[noitemsep,topsep=0pt]
    \item 1st tier grey literature (measure=1): High credibility: 
    % \\ 
    Books, magazines, theses, government reports, white papers
    % \\ 
    \item 2nd tier grey literature(measure=0.5): Moderate credibility: 
    % \\ 
    Annual reports, news articles, presentations, videos, 
    % \\ 
    Q/A sites (such as StackOverflow), Wiki articles
    % \\ 
    \item 3rd tier grey literature(measure=0): Low credibility: 
    % \\ 
    Blogs, emails, tweets
\end{itemize}

% \end{tabular} 
& Assign score based on type of the artifact.
\\ \hline
% \end{tabular}
\end{longtable}}
% \end{table}

\change{To specify the level of confidence, we combined two factors for each practice as shown in Table~\ref{confidence}: (1) The average quality of the artifacts that the practice is mentioned in; and (2) The number of artifacts that point to the practice. We defined four level of confidence: High, Moderate-High quantity, Moderate-High quality and Low. In Table~\ref{confidence}, the range for average quality score of artifacts is between 0 and 20. We divided this range to 2 equal width ranges. The number of artifacts analyzed in this step is 66, and we divide the range of 0 to 66 to 2 equal width ranges. The numbers in front of each level of confidence is the number of practices fell into the category. Moderate-High quantity is used when the practice is mentioned in more than a half of the artifacts but the average quality of the artifacts is lower than the selected threshold which is 10. Moderate-High quality is used when the practice is mentioned in less than a half of the artifacts but the average quality of the artifacts is more than the threshold. We will refer to this table in Section~\ref{practices}.}

\begin{table}[t]
{\color{black}
\centering
\caption{The level of confidence.}
    \label{confidence} 
\begin{tabular}{|l|l|l|}
\hline
                    & \multicolumn{2}{c|}{Average quality score of artifacts} \\ \hline
Number of artifacts & {[}10,20{]}    & {[}0,10)     \\ \hline
{[}33,66{]}         & High (1)          & Moderate-High quantity (0) \\ \hline
{[}0,33)           & Moderate-High quality (16)     & Low (0)          \\ \hline

\end{tabular}}
\end{table}

\subsection{Step Three: Searching Company-specific Grey Literature}

Some artifacts collected in Step One were company-specific artifacts. Additionally, some artifacts contained a list of companies that use feature toggles. From these artifacts, we obtained a list of companies which use feature toggles in their development cycle. Additional searches were conducted to collect more artifacts related to feature toggles from these specific companies. We used the search strings in the following format:
``[company name] [feature toggle term]'' 
where company name represents the name of the company; and feature toggle term is a search term for ``feature toggle,'' as defined in Step One. For each combination of company name and feature toggle term, a search string was applied to collect as many artifacts as possible. These strings were searched by using both the Google search engine and search feature found within a company's blog. We looked at the first 10 pages of the Google search engine result and all of the results in the company's blogs. If a company uses a feature toggle management system named by an artifact, we also used that system's name instead of ``feature toggle term'' in a search string. For example, Facebook uses Gatekeeper for feature toggle management \cite{feitelson2013development}. We used Gatekeeper instead of ``feature toggle term'' as well as search terms for feature toggle in the search for Facebook.

\subsection{Step Four: Extraction of Practice Usage from Company-specific Artifacts}

We analyzed the company-specific grey literature artifacts collected in Step One and Step Three to determine which practices identified in Step Two are used by the companies, as mentioned in the artifacts. If a practice was not clearly mentioned, a second person analyzed the artifact and then we made a decision if the company used the practice or not. %The practices used by each company were recorded in a spreadsheet containing list of companies and the list of identified practices.

Step Three and Step Four in Figure~\ref{fig:methodology} were performed iteratively and repeatedly if new artifacts for a company were found in Step Four.

\subsection{Step Five: Conducting Survey}

After extracting practice usage by the companies, we observed that our results were not complete. For instance, some of the identified feature toggle practices were not mentioned in any of the company-specific artifacts. \change{In addition, we were interested to know about the status of usage of feature toggles in industry.} So, we conducted a survey to obtain more information about feature toggle practice usage.

Contact information of company employees was gathered by collecting social media accounts and email addresses of named individuals associated with company-specific artifacts found in Steps One and Three. We also found contact information of managers/developers in companies that we knew were using feature toggles based upon Step One, even though we did not find company-specific artifacts for them in Step Three. We requested each practitioner to complete the survey. We contacted the practitioners by email where email was available and by social media if email addresses were not found. \change{We sent the survey to 45 companies and got 20 responses for a response rate to the survey of 44\%.}

The survey has 11 questions and is presented in the Appendix. On average, each practitioner needed approximately 5 minutes to answer all questions. We used Likert scale options \cite{likert1932technique} for the 12 practices for which Likert scale options can be used. We provided five options in the survey for each practice to specify how much the survey respondents use the practice: Always, Mostly, About half of the time, Rarely, and Never. For the remaining 5 practices, we provided practice-specific answer options. 

\subsection{Step Six: Analysis of Usage Frequency of Practices}

We analyzed the information from Step Four (analyzing company-specific artifacts) and Step Five (survey) to find the frequency of usage of each identified practice in the industry to answer RQ2. We integrated the result of Step Four and Step Five and report the frequency of usage of feature toggle practices. \change{In addition, we reviewed all the artifacts (including initial artifacts and company-specific artifacts) and record any comparison made between practices in artifacts in this step.}

\section{Results} \label{result}

\begin{figure}[t]
\centerline{\includegraphics[width=\textwidth]{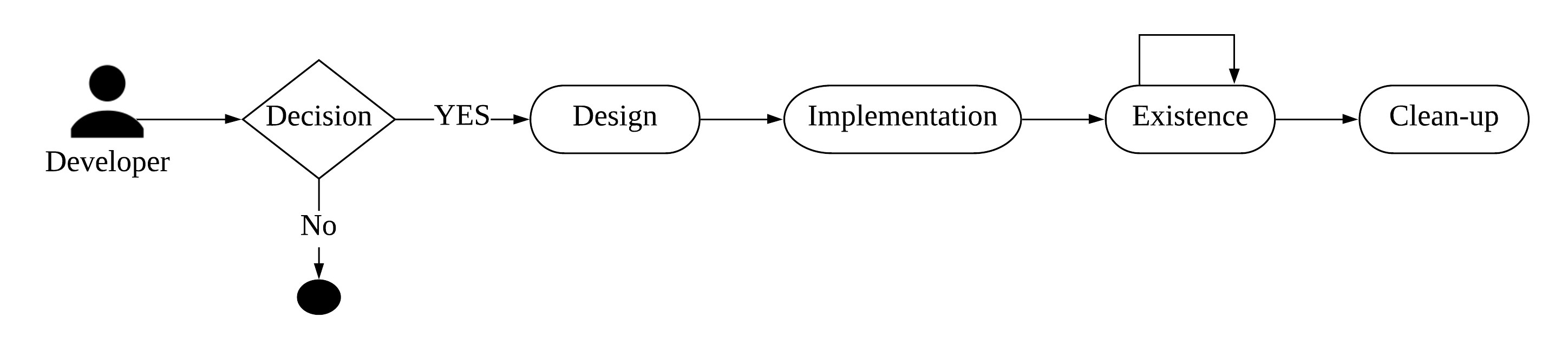}}
\caption{\change{The lifecycle of a feature toggle, adapted from \cite{tiwarifeaturetoggles}.}}
\label{fig:lifecycle}
\end{figure}

\change{Based on \cite{tiwarifeaturetoggles}, we propose the lifecycle of a feature toggle as shown in Figure \ref{fig:lifecycle}. The first phase is \textbf{Decision}, when the development team decides if the usage of a feature toggle is necessary for their situation. When the development team decides to use a feature toggle, the second phase is \textbf{Design} in which the details of the feature toggle is determined, such as type of the toggle, the possible values of the toggle, and/or the name of the toggle. The third phase is \textbf{Implementation} which includes adding the designed feature toggle to the code. The fourth phase is \textbf{Existence}, in which the toggle exist in the code and/or is updated. The fifth phase is \textbf{Clean-up} in which the toggle is removed from the code. For each identified practice, in Section \ref{practices}, we will specify the lifecycle phases covered by the practice.}

In the rest of this section, we present the results of the research methodology. Section \ref{practices} provides the answer to RQ1-Identification, and Section \ref{frequency} provides the answer to RQ2-Frequency. 

\subsection{Feature Toggles Practices} \label{practices}
We found 66 artifacts \change{including 10 peer-reviewed papers, 41 blog posts and online articles, and 15 videos} in Step One. From these artifacts, we identified and categorized 17 practices in Step Two. We found four categories of practices: Management practices (6), Initialization practices (3), Implementation practices (3), and Clean-up practices (5). We describe the 17 practices in their categories in the rest of this section. 

\change{The level of confidence of each identified practice is calculated as we described in Step Two of the methodology in Section \ref{methodology}. The range of scores of quality assessment of grey literature artifacts is from 9.5 to 17. The number of practices that fell into each level of confidence is shown in Table~\ref{confidence}. ``Use management systems'' is the only practice in \textbf{High} level, all the rest practices are in \textbf{Moderate-High quality} level.}

\change{Some of the practices are related to other practices. For example, following one of the practices may help following another practice. We determined the relation between practices based on the effects of using them. Related practices are summarized in Table~\ref{relatedpractices}. }

\change{For each practice, we explain the practice using the following structure:
\begin{enumerate}
    \item Description: The explanation of the practice. 
    \item Goal: The goal of following the practice.
    \item Examples: The example(s) from the 38 companies listed in Table~\ref{tab2} and practitioner's experiences about following or not following the practice.
    %\item Related practices: Some of the practices are related to other practices. For example, following one of the practices may help following another practice. Related practices are summarized in Table~\ref{relatedpractices}.
    \item Covered phases in lifecycle: The phase(s) in which the practice would be used, as shown in Figure~\ref{fig:lifecycle}.
    \item Comparative practices: The list of practices which could be compared with the practice (because of the same goal). Practices in Clean-up category are the only practices which could be compared to each other. So, practices in other categories do not have this bullet in their structure.
    %\item Level of confidence: The level of confidence of the artifacts, as we described in Step Two of the methodology in Section \ref{methodology}.
    \item Generalizability:  A categorization of whether the practice is general to software engineering (SE); common with configuration options (C) as identified in \cite{sayagh2018software} and discussed in Section~\ref{background-relatedworks}; or feature toggle-specific (F).
\end{enumerate}
}

%\subsubsection{Management practices (6)}
%\ovalbox{Management }
\begin{tcolorbox}[colback=black!5!white,colframe=black,size=small]
  Management Practices (6)
\end{tcolorbox}
Management practices are practices that are performed by development team members to make decisions about how to use feature toggles.
\begin{description}
\item{\textit{\textbf{M1: Use management systems}}: }

\change{\textbf{Description:}} Management systems help companies to create, use, and change the value of feature toggles. Using feature toggle management systems helps to overcome technical debt and manage the added complexity \cite{rolloutmanagementsystem}. Feature toggle management systems can have a dashboard that helps team members to see the list of feature toggles and their current values. Team members can add new feature toggles or change the values of the toggles if they have the permission. Management systems are connected to the code, and the changes impact the running system immediately. Management systems can be open-source or closed-source. Organizations may create their own feature toggle management system. 

\change{\textbf{Goal:} To manage creating and updating feature toggles in a centralized system.} %As was mentioned, adding feature toggles increases the complexity of the code, and managing a number of feature toggles could be challenging. 

\change{\textbf{Examples:}} Facebook uses the Gatekeeper toggle management system \cite{feitelson2013development}. Alternatively, companies can use third-party management systems, such as LaunchDarkly\footnote{\url{https://launchdarkly.com/}} and Split\footnote{\url{https://www.split.io/}}. As an example,   Behalf\footnote{\url{https://www.behalf.com/}} and CircleCI\footnote{\url{https://circleci.com/}} use the LaunchDarkly feature toggle management system. Envoy\footnote{\url{https://envoy.com/}} uses Split's feature toggle management system.

%\change{\textbf{Related practices:} Using management system could be related to the following practices based on the management system implementation: ``Document feature toggle's information'', ``Log changes'', ``Give access to team members'', ``Group the feature toggles'', ``Set up default values'', ``Determine the type of the toggle'', ``Type of the assigned values'', ``Ways  of  accessing  the  values'', ``Store  type'', ``Automatic reminders'', ``Track unused toggles'', ``Limit the number of feature toggles''. The management system can provide facility to follow each of the mentioned practices.}

\change{\textbf{Covered phases in lifecycle:} Design, Implementation, Existence, and Clean-up.}

%\change{\textbf{Comparative practices:} N/A.}

%\change{\textbf{Level of confidence:} High}

\change{\textbf{Generalizability:} (C). Sayagh et al. \cite{sayagh2018software} recommend the adoption of existing configuration frameworks.}
\\
\item{\textit{\change{\textbf{M2: Document feature toggle's metadata}}}: }

\change{\textbf{Description:}} Through the documentation of feature toggles, practitioners record the feature toggle's information e.g. the owner of the feature toggle; the current status (to remove untriaged, keep, removed); the time of its creation; and any notes. %Developers need to keep metadata on the \change{documents} consistent and up-to-date. Also, the management system can incorporate a \change{documentation's tool}.
%The metadata of feature toggles in the documentation are valuable for developers. For example, if the documentation include notes about the purpose of usage of feature toggles, the new team member can understand the goal of using each feature toggle.

\change{\textbf{Goal:} To enable practitioners to have access to the feature toggles' metadata at any time.}

\change{\textbf{Examples:}} Google has a spreadsheet with a list of feature toggles, the owner, toggles' status, and notes about toggles which used in the Chrome project \cite{chromiumflags}. \change{However, this spreadsheet has only changed three times in 2018 and three times in 2019. The developers may have moved on to use a new tool for documenting feature toggles' metadata.}

%\change{\textbf{Related practices:} ``Use management systems''.}

\change{\textbf{Covered phases in lifecycle:} Implementation, Existence, and Clean-up}.

%\change{\textbf{Comparative practices:} N/A.}

%\change{\textbf{Level of confidence:} Moderate.}

\change{\textbf{Generalizability:} (SE\cite{stephens2015beginning}/C). Sayagh et al.  \cite{sayagh2018software} listed ``Comprehension of Options and Values'' and ``Configuration Knowledge Sharing'' as two configuration option activities . The concept behind these two activities is similar to documenting feature toggle's metadata.}
\\
\item{\textit{\textbf{M3: Log changes}}:}

\change{\textbf{Description:}} Tracking changes that are made on feature toggles. By logging, the information of who changes which toggle and when is recorded \cite{launchdarklybestpractices}.

\change{\textbf{Goal:}} To document traceability of actions for creating, updating, and deleting toggles and their values.

\change{\textbf{Examples:}} Split's feature toggle management system has the ability to log changes of the feature toggles \cite{splitauditlogs}. 

%\change{\textbf{Related practices:} Logging changes of feature toggles is related to the following practices: 1) ``Use management systems'', 2) ``Track unused toggles'': if the changes on feature toggles are logged, then finding unused toggle is easy for developers.}

\change{\textbf{Covered phases in lifecycle:} Implementation, Existence, and Clean-up.}

%\change{\textbf{Comparative practices:} N/A.}

\change{\textbf{Level of confidence:} Moderate.}

\change{\textbf{Generalizability:} (SE \cite{stephens2015beginning}/C). In \cite{sayagh2018software} having ``Right Granularity of Execution Logs'' is one of the recommendations, so logging changes is recommended for configuration options.} 
\\ 
\item{\textit{\textbf{M4: Determine applicability of feature toggle}}:}

\change{\textbf{Description:}} Before the design and implementation of a feature toggle, the development team should determine if a feature toggle should be used. Using feature toggles adds more decision points to the code which adds more complexity to the code and requires attention to remove toggles when the initial use is completed. 

\change{\textbf{Goal:}} To make an explicit decision on the creation of a toggle which may reduce the number of feature toggles in a code base. %Feature toggles are useful, but they are not essential for some situations. 
%For instance, a developer can integrate the functionality into the trunk branch of a product without a user interface (UI) element such that the partially-developed code can be tested through the application programming interface (API) without the feature being accessible to a user. When the feature is completed, the UI element could be added and rolled out to the user instead of initially adding the UI element and wrapping it in a feature toggle \cite{fowlerfeaturetoggle}. 

\change{\textbf{Examples:}} Different companies have different approaches to making decisions. For example, all new features in GoPro have feature toggles\footnote{\label{gopro} https://bit.ly/2ISi1ye}. However, practitioners in Finn.no, a largest online marketplace in Norway, avoid using feature toggles if they do not need the toggle \cite{finnunleash}.

%\change{\textbf{Related practices:} This practice has no related practice.} 

\change{\textbf{Covered phases in lifecycle:} Decision.}

%\change{\textbf{Comparative practices:} N/A.}

%\change{\textbf{Level of confidence:} Moderate.}

\change{\textbf{Generalizability:} (F). }%Because of the various types of the toggles, various goals of using them and various management and implementation of them it is important to determine the applicability of the toggles.
\\
\item{\textit{\textbf{M5: Give access to team members}}:}

\change{\textbf{Description:}} Through this practice, permission to change values of feature toggles is granted to team members in addition to developers using the feature toggle management system.

\change{\textbf{Goal:}} To prevent feature toggle management bottleneck.

\change{\textbf{Examples:}} If all team members, such as Q\&A team members, have access to feature toggles, they can change a toggle status in case of a problem \cite{launchdarklybestpractices}. For instance, Instagram gives access to their feature management system to the product managers and sales team \cite{instagramflexible}.

%\change{\textbf{Related practices:} ``Use management systems''.}

\change{\textbf{Covered phases in lifecycle:} Design, Implementation, and Existence.}

%\change{\textbf{Comparative practices:} N/A.}

%\change{\textbf{Level of confidence:} Moderate.}

\change{\textbf{Generalizability:} (SE).}
\\
\item{\textit{\textbf{M6: Group the feature toggles}}:}

\change{\textbf{Description:}} \change{Grouping similar feature toggles.}

\change{\textbf{Goal:}} To enable the assignment of access to groups of feature toggles to teams or team members \cite{launchdarklybestpractices}; to simplify management of dependent toggles \cite{tiwarifeaturetoggles}; or to turn on or off feature toggles related to part of the system at the same time.

\change{\textbf{Examples:}} Practitioners in GoPro have a two-level toggle hierarchy: simple feature toggles and higher level feature toggles\cite{circlciwebinar}. 

%\change{\textbf{Related practices:} ``Use management systems''.}

\change{\textbf{Covered phases in lifecycle:} Design, Implementation, and Existence.}

%\change{\textbf{Comparative practices:} N/A.}

%\change{\textbf{Level of confidence:} Moderate.}

\change{\textbf{Generalizability:} (SE).}
   
\end{description}

%\subsubsection{Initialization practices (3)}
\begin{tcolorbox}[colback=black!5!white,colframe=black,size=small]%hbox
  Initialization Practices (3)
\end{tcolorbox}
Initialization practices are used to make decisions about the design of the feature toggle before their creation.
\begin{description}
\item{\textit{\textbf{I1: Set up the default values}}:}

\change{\textbf{Description:}} Default values for each feature toggle are set in case if assigned values are not found or do not exist.

\change{\textbf{Goal:}} To mitigate unwanted behavior of the feature toggle. 

\change{\textbf{Examples:}} At Lyris, feature toggles without values are automatically turned off \cite{sowafeaturebits}.

%\change{\textbf{Related practices:} ``Use management systems''.}

\change{\textbf{Covered phases in lifecycle:} Design and Implementation.}

%\change{\textbf{Comparative practices:} N/A.}

%\change{\textbf{Level of confidence:} Moderate.}

\change{\textbf{Generalizability:} (SE/C). Sayagh et al.  \cite{sayagh2018software} mentioned that the selection of the ``right'' default value for configuration options is a challenge.}
\\
\item{\textit{\textbf{I2: Use naming convention}}:}

\change{\textbf{Description:}} Having predefined naming rules for feature toggles. 

\change{\textbf{Goal:}} To establish naming conventions, particularly to make the intention of the toggle self-documented.

\change{\textbf{Examples:}} Having the naming convention has several benefits. First, understanding the purpose of using the toggle is useful \cite{sowafeaturebits}, i.e. if the owner of the code is changed, the new owner can understand the usage of the toggle easily if the name of the toggle reflects its usage. An example of Lyris toggles is ``ct.enable\_flex\_cache\_inspector' and the purpose of using the toggle is clear based on its name \cite{sowafeaturebits}. Second, the use of a naming convention reduces the likelihood of  multiple toggles with same names even by different teams by following naming conventions \cite{launchdarklybestpractices}. Third, adding the type of the toggle as a prefix in its name can help with the management of the toggles \cite{ulnodevops}. For instance, if the feature toggle is a short-lived toggle, like release toggles, the developer will get a signal from the name of the toggle that the first intention of using the toggle was a short-term use and will plan to remove it. For instance, In InVision, long-lived toggles have ``OPERATIONS-'' prefix \cite{bennadaljson}. Developers in this company also add the JIRA ticket number to the name of the feature toggle to make the purpose of using the toggle and responsible team to remove the toggle clear. If a ``RAIN-123-release-the-kraken'' is a name of the toggle, it is clear that the toggle is related to JIRA ticket RAIN-123 and the responsible team to clean-up the feature toggle is the Rainbow team \cite{invisionlaunchdarkly}. 

%\change{\textbf{Related practices:} This practice has no directly related practice. However, the type of the toggle can reflects in the name of the toggle and this is related to ``Determine the type of the toggle'' practice,Such as the InVision example in the ``Effects and Examples'' above.}

\change{\textbf{Covered phases in lifecycle:} Design and Implementation}.

%\change{\textbf{Comparative practices:} N/A.}

%\change{\textbf{Level of confidence:} Moderate.}

\change{\textbf{Generalizability:} (SE\cite{dale2003introduction}/C). In \cite{sayagh2018software} one of the challenges of using configuration options is ``meaningless Option Names'' .The authors listed ``Explicit Option Naming Convention'' as a recommendation to overcome the challenge which is similar to use naming convention practice.}
\\
\item{\textit{\textbf{I3: Determine the type of the toggle}}:}

\change{\textbf{Description:}} With this practice, the type of the toggle is specified using the toggle types mentioned in Section \ref{background-background}. The implementation and management of each type of the four toggle types are different.

\change{\textbf{Goal:}} To aid in quality management of a toggle's implementation and to enable the plan to remove the toggle on time based on the type of the toggle.

\change{\textbf{Examples:} The author in \cite{featureflagio} points to naming short-lived toggles with the prefix of ``temp-'' in their name. The identification of short-lived toggles can be useful in limiting the number of toggles \cite{hodgsonleanproduct}.} 

%\change{\textbf{Related practices:} This practice is related to the two following practices: 1)``Use management systems'', and 2) ``Limit the number of feature toggles'': limiting the number of feature toggles is a clean-up practice and is explained in the following sections. The relation between these  two practices is explained in the related practices of ``Limit the number of feature toggles''.}

\change{\textbf{Covered phases in lifecycle:} Design and Implementation.}

%\change{\textbf{Comparative practices:} N/A.}

%\change{\textbf{Level of confidence:} Moderate.}

\change{\textbf{Generalizability:} (F).}
\end{description}

%\subsubsection{Implementation practices (3)}
\begin{tcolorbox}[colback=black!5!white,colframe=black,size=small]
  Implementation Practices (3)
\end{tcolorbox}

Implementation practices are related to implementation details of feature toggles.
\begin{description}
\item{\textit{\textbf{Im1: Type of assigned values}}: }

\change{\textbf{Description:}} Companies use three different ways to assign values to toggles. One way is to assign a string to feature toggles. The second way to assign values is to assign Boolean values to feature toggles. When the value is true, the toggle is enabled \cite{meyerfeatureflags}. The third way is to assign multivariate values, such as when the toggle capture user experiences.

\change{\textbf{Goal:} To help practitioners to choose an appropriate type of the values for feature toggles in their system.} 

\change{\textbf{Examples:}} For instance, one of the feature toggles in Google Chrome project is kDisableFlash3d[] = ``disable-flash-3d''. If the value of the feature toggle is set then the toggle is enabled  \cite{rahman2016feature}.  As another example, Rollout\footnote{https://rollout.io/} provides multivariate toggles, for instance a toggle can accept ``Red'', ``Blue'' and ``Yellow'' as its value.

%\change{\textbf{Related practices:} ``use management systems''.}

\change{\textbf{Covered phases in lifecycle:} Design and Implementation.}

%\change{\textbf{Comparative practices:} N/A.}

%\change{\textbf{Level of confidence:} Moderate.}

\change{\textbf{Generalizability:} (C). Sayagh et al.  \cite{sayagh2018software} provide the ``Using Simple Option Types'' recommendation which mentioned string and Boolean types for configuration options. This recommendation has partially overlapped with the current practice.}
\\
\item{\textit{\textbf{Im2: Ways of accessing the values}: }}

\change{\textbf{Description:}} We identified three ways development teams access the values of feature toggles. First, the feature toggles could be primitive variables, hard-coded into the program.
Second, toggles could be objects and the object has a method to determine the value of the toggle (e.g. myToggle.isActive()). Third, toggles could be accessed through a manager object. Managers map key/value pairs to return the value.

\change{\textbf{Goal:} To help practitioners to choose an appropriate way of accessing the values in their system.}

\change{\textbf{Examples:} } Figure \ref{fig:toggle} is an example of directly access to the values. We found implemented libraries in GitHub which use a method from toggle object, such as rollout\footnote{https://github.com/fetlife/rollout}. LaunchDarkly is an example of using manager objects to access feature toggle values.

\change{Based on the experience of the practitioner in \cite{danpiessensfeaturetoggle}, having class of toggles and check the value of the toggle using isEnable() function of the class is better than checking the primitive variable of a string name of the toggle. Having feature toggle objects helps to refactor toggles same as the other parts of the code and find every usage of it easily. The practitioner in \cite{benjamindaydevops} also points to the fact that using objects of feature toggles is better than strings because of getting compile error on all places a feature toggle is used after removing the toggle. The same comparison is mentioned in \cite{robertsfeaturetoggle}: ``Toggles should be real things (objects) not just a loosely typed string. This helps with removing the toggle after use: 1) Can perform a `find uses' of the Toggle class to see where it's used, and 2) Can just delete the Toggle class and see where build fails.''} 

%\change{\textbf{Related practices:}``Use of management systems''.}

\change{\textbf{Covered phases in lifecycle:} Design and Implementation.}

%\change{\textbf{Comparative practices:} N/A.}

%\change{\textbf{Level of confidence:} Moderate.}

\change{\textbf{Generalizability:} (SE \cite{dale2003introduction}).}
\\
\item{\textit{\textbf{Im3: Store type}}:}

\change{\textbf{Description:}} The list of feature toggles and their values can be stored in one of two ways: file storage and database storage. In file storage, the values of feature toggles are stored in one or multiple configuration files%, such as what Google does in the Chrome project
\cite{rahman2016feature}. In database storage, the values of feature toggles are stored in databases, such as Redis \cite{meyerfeatureflags} or SQL \cite{robertsfeaturetoggle}. In addition, some companies ``use a third party service'' to fetch values of the feature toggle. If they use a feature management system, they fetch the values of feature toggles from the management system.

%Dropbox uses both configuration files and database. A JSON file called stormcrow\_config.json is shared between all the production servers and contains the value of feature toggles. If this JSON file is not found for any reason, the feature toggle management system Stormcrow has the ability to access to the database directly \cite{dropboxstormcrow}

\change{\textbf{Goals:} To store the values of feature toggle in an appropriate way, based on the advantages and disadvantages of each store type.}

\change{\textbf{Examples:} Based on the article \cite{hodgsonfeaturetoggle}, when practitioners use configuration files to save the feature toggle values, they may need to redeploy the application after each value update to get the reflection of the updated value. In addition, when a system is in large scale, it is hard to manage feature toggles using configuration files and it is hard to make sure the consistency of configuration files on different servers. So the practitioner recommend to use some sort of database to store the value of feature toggles. } 

\change{The practitioner in \cite{meyerfeatureflags} is also make the same comparison in his article: `` A more dynamic approach is to store the feature configuration in either an ephemeral or permanent storage, like Redis or your database, respectively.
Assuming your code continuously checks feature flags at runtime, all it takes is changing a value in the central configuration service and it can have an immediate affect on the running application without requiring a restart.''}

\change{The article \cite{surfingthecode} points to the same comparison: `` Release and Experiment toggles are likely to be set at deployment time, so from a running application perspective they are \textit{static} settings... However, Ops and Permission toggles are \textit{dynamic} and need to be configurable at run time, so you might want to store them in a database of some sort.''  The same concept is mentioned in \cite{tiwarifeaturetoggles} and \cite{benjamindaydevops} as well.} 

\change{Dropbox use configuration files and database together\cite{dropboxstormcrow}. They explain the reason of using both ways. Because of having a large number of production servers, they prefer using a database instead of using configuration files. However, this may create a huge number of fetches against the database and the database will be the single point of the failure even if they have caching system. So, they come up with a combination of both ways. A JSON file is shared between all the production servers and  contains  the  value  of  feature  toggles. If  this  JSON  file  is  not  accessible for any reason, the feature toggle management system has the ability to access to the database directly. This example shows the advantages and disadvantages of using each way of storing the values of feature toggles.}
%\change{\textbf{Related practices:} ``Use management practices''.}

\change{\textbf{Covered phases in lifecycle:} Design and Implementation.}

%\change{\textbf{Comparative practices:} N/A.}

%\change{\textbf{Level of confidence:} Moderate.}

\change{\textbf{Generalizability:} (C). ``Managing Storage Medium'' activity form \cite{sayagh2018software} is similar to this practice.  Moreover, one of the challenges in \cite{sayagh2018software} is ``Storage media Impact Performance'' which is really similar to the practitioners' experiences we provided for this practice.}

\end{description}

%\subsubsection{Clean-up practices (5)}
\begin{tcolorbox}[colback=black!5!white,colframe=black,size=small]
  Clean-up Practices (5)
\end{tcolorbox}

Following the clean-up practices helps practitioners to remove their feature toggles on time and manage the complexity of using feature toggles.

\begin{description}
\item{\textit{\textbf{C1: Add expiration date}}: }
\change{This practice is followed using one of the following three processes:}
\begin{itemize}
    \item C1.1: Time bombs: 
    
    \change{\textbf{Description:}} If the toggle exists after its expiration date, a test fails or the application does not start, which causes a developer to remove the toggle \cite{hodgsonfeaturetoggle}, \cite{leenamergehells}, \cite{hodgsonleanproduct}. The expiration date is the latest possible date which the developers should remove the toggle from the code. Using this practice forces practitioners to remove a toggle by the determined expiration date.
    
    \change{\textbf{Goal:} To remove unused toggles.} 
    
    \change{\textbf{Examples:} We did not find any specific examples in company-specific artifacts. However, the practitioner in \cite{hodgsonleanproduct} says: ``(Time bomb) is very extreme and I wouldn't recommend doing (it). I think a lot of organizations would not be comfortable with doing that, it does force a lot of other things to be good.''}
    
    %\change{\textbf{Related practices:} The practice has no related practices.}
    
    \change{\textbf{Covered phases in lifecycle:} Design, Implementation, and Clean-up. The expiration date is identified during the design phase, the time bomb is added in the Implementation phase and the feature toggle is removed in the Clean-up phase.}
    
    \change{\textbf{Comparative practices:} All the practices in this category (C1-C5).}
    
    %\change{\textbf{Level of confidence:} Moderate.} 
    
    \change{\textbf{Generalizability:} (F).}
    \\
    \item C1.2: Automatic reminders: 
    
    \change{\textbf{Description:}} Add automatic reminders to remind developers the deadline for removing feature toggles \cite{leenamergehells}. Using this practice helps practitioners to remember to remove a toggle by the determined deadline.
    
    \change{\textbf{Goal:} To remove unused toggles.} 
    
    \change{\textbf{Examples:}} Slack has an archival system. When developers want to add a new feature toggle, they have to specify the date they plan to delete the toggle. If the toggle is not deleted by the specified date, the developer will get an alert\footnote{https://bit.ly/2W4hQUk}.
    
    %\change{\textbf{Related practices:} The automatic reminders can be implemented in management systems, so the practice is related to ``Use management systems'' practice.}
    
    \change{\textbf{Covered phases in lifecycle:} Design, Implementation, and Clean-up.}% The removing deadline is identified during the design phase, the reminder is added in the Implementation phase and the feature toggle is removed in the Clean-up phase.
    
    \change{\textbf{Comparative practices:} All the practices in this category (C1-C5).}
    
    %\change{\textbf{Level of confidence:} Moderate.}
    
    \change{\textbf{Generalizability:} (SE).}
    \\
    \item C1.3: Use cards/tasks/stories for removing toggles:
    
    \change{\textbf{Description:}} Add tasks/stories/cards for removing toggles to a Kanban board (or any other tool that the team uses) \cite{leenamergehells} or to developer's task backlog \cite{hodgsonfeaturetoggle}, \cite{hodgsonleanproduct}. Using this practice reminds practitioners the task of removing a toggle at the expiration date when the purpose of using the toggle is done.
    
    \change{\textbf{Effects:} To remove unused toggles.} 
    
    \change{\textbf{Examples:}} Developers at Lyris create user stories for removing toggles \cite{sowafeaturebits}. \change{However, the practitioner in \cite{hodgsonleanproduct} says about his experience of using this practice:`` I think it  can help but it's kind of the bare minimum, .. I have spent a lot of time with clients where the clean-up ticket is just at the very top of the next Sprint's backlog for like six months. It's always like yeah we really should do that the next week and it will be done next week, but it's always next week.'' Practitioner in \cite{invisionlaunchdarkly} has the same experience with this technique: `` Teams are supposed to create additional ``clean-up'' tasks in JIRA for their feature flags such that we don't lose track of them. The reality, however, is far less sanitary. Our feature flags tend to pile up and we have to occasionally have a ``purge'' of flags that no longer seem relevant.''} 
    
    %\change{\textbf{Related practices:} The practice has no related practices.}
    
    \change{\textbf{Covered phases in lifecycle:} Design, Implementation, and Clean-up.}
    
    \change{\textbf{Comparative practices:} All the practices in this category (C1-C5).}
    
    %\change{\textbf{Level of confidence:} Moderate.} 
    
    \change{\textbf{Generalizability:} (SE \cite{hammarberg2014kanban}).}
\end{itemize}

\item{\textit{\textbf{C2: Track unused toggles}}:}

\change{\textbf{Description:}} With this practice, dead code and unused feature toggles are removed. Based on the logging system or using documentation, the status of toggles could be monitored. Developers can use this data to find when the toggle is safe to remove \cite{launchdarklybestpractices}. When a toggle is always on or always off, it should be removed.

\change{\textbf{Goal:} To remove unused toggles.} 

\change{\textbf{Examples:} DropBox has a static analyzer tool with a service specifically for feature toggles. The static analyzer tool sends emails to feature toggle owners about removing the toggles which are not in use anymore \cite{dropboxstormcrow}.}

%\change{\textbf{Related practices:} ``Use management systems'', ``Document feature toggle's metadata'' and ``Log changes''.}

\change{\textbf{Covered phases in lifecycle:} Implementation and Clean-up.}

\change{\textbf{Comparative practices:} All the practices in this category (C1-C5).}

%\change{\textbf{Level of confidence:} Moderate.}

\change{\textbf{Generalizability:} (F).}
\\
\item{\textit{\textbf{C3: Limit the number of feature toggles}}:}

\change{\textbf{Description:}} Using this practice the number of alive feature toggles at a time are limited to control the number of toggles. An alive feature toggle is a toggle which exists in the code whether it is on or off. By this limitation, practitioners have to remove an unused toggle to be able to add a new toggle if the number of existing toggles meets the limitation \cite{hodgsonfeaturetoggle}, \cite{hodgsonleanproduct}, \cite{schermann2018we}. 

\change{\textbf{Goal:} To remove unused toggles.} 

\change{\textbf{Examples:} We did not find any specific example of using this practice in company-specific artifacts. The practitioner in \cite{hodgsonleanproduct} says this practice is his favorite practice for removing feature toggles.}%``This is like my favorite technique for reducing the number of Flags. It is just artificially saying we'll only have so many flags because it aligns the incentives of the different people working on the system.''

%\change{\textbf{Related practices:} If the type of the toggles are pre-determined using ``Determine the type of the toggle'', the practitioners have a list of short-lived toggles as a suggested list of toggles to remove. Instead of checking all of the toggles, the short-lived toggles could be checked for removal.} 

\change{\textbf{Covered phases in lifecycle:} Implementation and Clean-up}.

\change{\textbf{Comparative practices:} All the practices in this category (C1-C5).}

%\change{\textbf{Level of confidence:} Moderate.}

\change{\textbf{Generalizability:} (F). Sayagh et al.  \cite{sayagh2018software} points to minimizing the number of configuration options in the system as one the recommendations. This recommendation is partially covered the practice of limiting the number of feature toggles. In limiting the feature toggles, developers have to remove a toggle if the limitation is reached but removing is not mentioned in the recommendation of configuration options. So, we specify the practice as feature toggle specific practice.}
\\
\item{\textit{\textbf{C4: Create a cleanup branch}}:}

\change{\textbf{Description:}} This is the practice of creating a branch to delete the toggle and submitting a pull request for the branch at the same time as adding a new feature toggle \cite{ulnodevops}, \cite{featureflagio}. Using this practice prevents to forget the deletion of the feature toggle.

\change{\textbf{Goal:} To remove unused toggles.} 

\change{\textbf{Examples:} We did not find any example of using this practice in company-specific artifacts. The author of the \cite{featureflagio} says that this practice works pretty well in their team: ``The advantage to managing cleanup this way is that you do the work to remove the flag when all of the context is fresh in your mind. At this point, you know all the pieces that get touched by the change, and it is easier to be sure you don't forget something .. %You will need to merge master back into your cleanup branch periodically, but that is usually easier than it would be to recall all of the context relating to the original change.
This is certainly not the only way to handle this issue, but it seems to work pretty well for our team.''.}

%\change{\textbf{Related practices:} The practice has no related practices.}

\change{\textbf{Covered phases in lifecycle:} Implementation and Clean-up.}

\change{\textbf{Comparative practices:} All the practices in this category (C1-C5).}

%\change{\textbf{Level of confidence:} Moderate.}

\change{\textbf{Generalizability:} (SE).}
\\
\item{\textit{\textbf{C5: Change a feature toggle to a configuration setting}}: }

\change{\textbf{Description:}} This is the practice of keeping feature toggles in the code with changed functionality. The feature toggle can be changed to admin or user configuration settings. This technique is used when development team decide to keep more than one variant of the feature toggle in the code.

\change{\textbf{Goal:} To remove unused feature toggles.}

\change{\textbf{Examples:}} Suppose a feature toggle is used for running experiments to see which color is better for the ``buy'' button in an e-commerce application. The experimental results show that the users are happiest when they can control the color of the button. Instead of deleting the feature toggle, it will be changed to a user configuration setting \cite{rolloutretirement}.

%\change{\textbf{Related practices:} This practice does not have any related practices.}

\change{\textbf{Covered phases in lifecycle:} Implementation and Clean-up (C1-C5).}

\change{\textbf{Comparative practices:} All the practices in this category.}

%\change{\textbf{Level of confidence:} Moderate.}

\change{\textbf{Generalizability:} (F).}
\end{description}

\begin{table*}[btp] 
{\color{black}
\setlength{\tabcolsep}{3pt}
\caption{Related practices.}
\scriptsize
\begin{tabular}{l p{2.5cm} *{17}{|c}|}
\toprule
\multicolumn{2}{c|}{ } & 
\multicolumn{6}{c|}{Management} &
\multicolumn{3}{c|}{Initialization} &
\multicolumn{3}{c|}{Implementation} &
\multicolumn{5}{c}{Clean-up} \\ %\cline{3-19}
   &
   &
  \rotatebox[origin=r]{90}{Use management systems} &
  \rotatebox[origin=r]{90}{Document feature toggle's metadata} &
  \rotatebox[origin=r]{90}{Log changes} &
  \rotatebox[origin=r]{90}{Determine  applicability  of feature toggle} &
  \rotatebox[origin=r]{90}{Give access to team members} &
  \rotatebox[origin=r]{90}{Group the feature toggles} &
  \rotatebox[origin=r]{90}{Set up the default values} &
  \rotatebox[origin=r]{90}{Use naming convention} &
  \rotatebox[origin=r]{90}{Determine  the  type  of  the toggle} &
  \rotatebox[origin=r]{90}{Type   of   assigned   values} &
  \rotatebox[origin=r]{90}{Ways  of  accessing  the  values} &
  \rotatebox[origin=r]{90}{Store  type} &
  \rotatebox[origin=r]{90}{Add expiration date} &
  \rotatebox[origin=r]{90}{Track unused toggles} &
  \rotatebox[origin=r]{90}{Limit  the  number  of  feature toggles} &
  \rotatebox[origin=r]{90}{Create a cleanup branch} &
  \rotatebox[origin=r]{90}{Change a feature toggle to a configuration setting} \\
%   \hline
  \midrule
\multirow{6}*{\rotatebox[origin=c]{90}{Management}} & Use management systems & -- & \checkmark & \checkmark & & \checkmark & \checkmark & \checkmark & & \checkmark & \checkmark & \checkmark & \checkmark & \checkmark & \checkmark & \checkmark & & \\ \cline{2-19}
& Document feature toggle's metadata & \checkmark  &  -- & & & & & & & & & & & & & & & \\ \cline{2-19}
& Log changes & \checkmark & & -- & & & & & & & & & &  & \checkmark & & & \\ \cline{2-19}
& Determine  applicability  of  feature  toggle & & & & -- & & & & & & & & & & & & &  \\ \cline{2-19}
& Give  access  to  team  members & \checkmark & & & & -- & & & & & & & & & & & &  \\  \cline{2-19}
& Group the feature toggles & \checkmark & & & & & -- & & & & & & & & & & & \\ 
% \hline
  \midrule

\multirow{3}*{\rotatebox[origin=c]{90}{Initialization}} & Set up the default values \rule[-8pt]{0pt}{20pt}  & \checkmark & & & & & & -- & & & & & & & & & &  \\ \cline{2-19}
& Use  naming  convention \rule[-8pt]{0pt}{20pt} & & & & \checkmark & & & & -- & & & & & & & & & \\ \cline{2-19}
& Determine the type of the toggle & \checkmark & & & & & & & & -- & & & & & & \checkmark & &  \\ 
% \hline
  \midrule

\multirow{3}*{\rotatebox[origin=c]{90}{Implementation}} & Type  of  assigned  values \rule[-8pt]{0pt}{20pt} & \checkmark & & & & & & & & & -- & & & & & & &   \\ \cline{2-19}
& Ways  of  accessing  the  values \rule[-8pt]{0pt}{20pt} & \checkmark & & & & & & & & & & -- & & & & & &  \\ \cline{2-19}
& Store  type \rule[-8pt]{0pt}{20pt} & \checkmark  & & & & & & & & & & & -- & & & & &  \\ 
% \hline
  \midrule

\multirow{5}*{\rotatebox[origin=c]{90}{Clean-up}} & Add expiration date  & \checkmark & & & & & & & & & & & & -- & & & &   \\ \cline{2-19}
& Track  unused  toggles & \checkmark & \checkmark & \checkmark & & & & & & & & & & & -- & & &   \\ \cline{2-19}
& Limit  the  number  of  feature  toggles & & & & & & & & & \checkmark & & & & & & -- & &   \\ \cline{2-19}
& Create  a  cleanup  branch & & & & & & & & & & & & & & & & -- &   \\ \cline{2-19}
& Change  a  feature  toggle  to  a  configuration  setting & & & & & & & & & & & & & & & & & -- \\ 
% \hline
  \bottomrule

\end{tabular}}
\label{relatedpractices}
\end{table*}

\subsection{Usage of Practices in Industry} \label{frequency}
\begin{figure}[t]
\centering
\includegraphics[width=250pt]{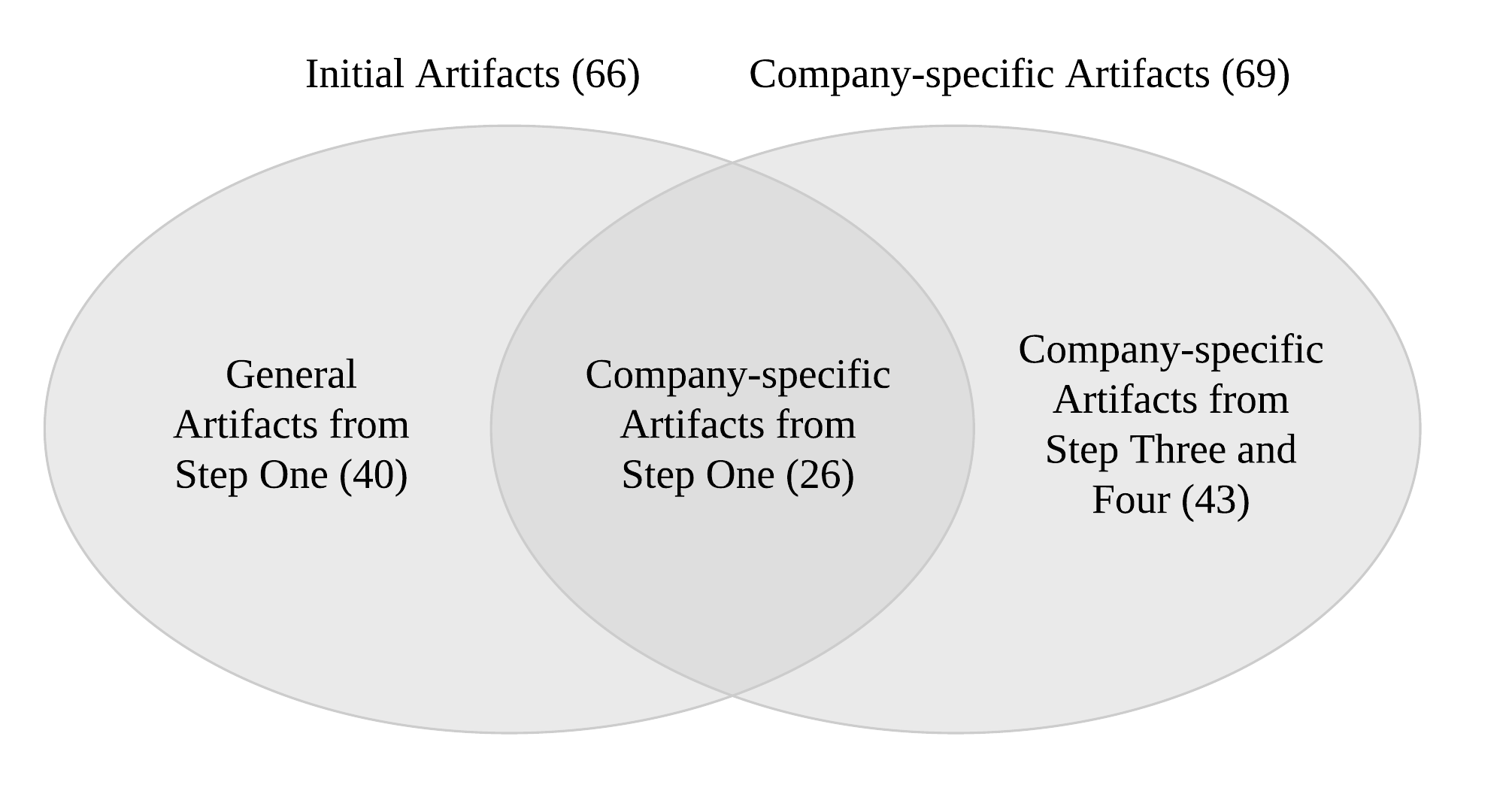}
\caption{Number of artifacts found in in Steps 1, 3 and 4}
\label{fig:venn}
\end{figure}

In Step One, 26 artifacts were company-specific artifacts.  In Step Three and Step Four, we found 43 additional company-specific artifacts. In total, 69 company-specific artifacts from 38 companies were collected. The overlap between initial artifacts and company-specific artifacts is shown in Figure~\ref{fig:venn}. In Step Four, we analyzed these 69 company-specific artifacts to find which companies use the identified practices. The practices used by each company are shown in Table~\ref{tab2} in the Appendix. This table is useful for practitioners because software practitioners prefer to learn through the experiences of other software practitioners \cite{moore2009crossing}, as we mentioned before.

\begin{table*}[btp] 
\caption{Feature toggle practices and their usage in industry}
% \begin{tabular}{p{102pt}|p{100pt}p{50pt}p{50pt}}
\begin{tabular}{p{3.6cm}p{2.5cm}p{50pt}p{50pt}}
% \hline
\toprule
Category (mean of frequencies based on artifacts, mean of frequencies based on survey) & Practice & Frequency from artifacts (38 companies)& Frequency from survey (20 companies) \\ 
% \hline
% \hline
\midrule
\multirow{6}*{{Management (42\%, 63\%)}} & Use management systems & 32 (84\%) & 20 (100\%) \\ \cline{2-4}
& Document feature toggle's metadata & 25 (66\%) & 7 (35\%) \\ \cline{2-4}
& Log changes & 21 (55\%) & 12 (60\%) \\ \cline{2-4}
& Determine  applicability  of  feature  toggle & 8 (21\%) & 16 (80\%) \\ \cline{2-4}
& Give  access  to  team  members & 7 (18\%) & 14 (70\%) \\  \cline{2-4}
& Group the feature toggles & 2 (5\%) & 7 (35\%) \\ 
% \hline
\midrule
\multirow{3}*{{Initialization (25\%, 72\%)}} & Set up the default values & 22 (58\%) & 17 (85\%) \\ \cline{2-4}
& Use  naming  convention & 5 (13\%) & 14 (70\%) \\ \cline{2-4}
& Determine the type of the toggle & 1 (3\%) & 12 (60\%) \\ 
% \hline
\midrule
\multirow{3}*{{Implementation (66\%, 100\%)}} & Type  of  assigned  values (string, boolean, multivariate, more than one) & 32 (1 (3\%), 7 (18\%), 5 (13\%), 19 (50\%)) & 20 (1 (5\%), 6 (30\%), 2 (10\%), 11 (55\%)) \\ \cline{2-4}
& Ways  of  accessing  the  values (primitive variable, objects, managers, more than one) & 28 (0 (0\%), 0 (0\%), 28 (74\%), 0 (0\%)) & 20 (3 (15\%), 5 (25\%), 3 (15\%), 9 (45\%)) \\ \cline{2-4}
& Store  type (file, database, both, third party service) & 15 (9 (24\%), 4 (11\%), 2 (5\%), - ) & 20 (3 (15\%), 6 (30\%), 6 (30\%), 4 (20\%)) \\ 
% \hline
\midrule
\multirow{5}*{{Clean-up (3\%, 39\%)}} & Add expiration date (Time  bombs, Automatic  reminders, Use   cards/tasks/stories   for   removing   toggles) & 6 (0 (0\%), 1 (3\%), 5 (13\%)) & 14 (1 (5\%), 4 (20\%), 9 (45\%)) \\ \cline{2-4}
& Track  unused  toggles & 1 (2\%) & 9 (45\%) \\ \cline{2-4}
& Limit  the  number  of  feature  toggles & 0 (0\%) & 10 (50\%) \\ \cline{2-4}
& Create  a  cleanup  branch & 0 (0\%) & 4 (20\%) \\ \cline{2-4}
& Change  a  feature  toggle  to  a  configuration  setting & 0 (0\%) & 2 (10\%)\\ 
% \hline
\bottomrule
\end{tabular}

\label{tab1}
\end{table*}

In Step Five, we conducted a survey to gather additional information about the usage of feature toggles practices in industry. We had company-specific artifacts of 38 companies. Of these 38 companies, we sent out the survey to 36 companies for which we had the contact information for release engineers and/or developers. In addition to these companies, we identified a list of 20 companies which use feature toggles in their companies. These companies are mentioned in artifacts as the example of companies which are using feature toggles but we cannot find company-specific artifacts related to their practice usage. We found contact information for release engineers and/or developers in 9 of these companies and sent the link of the survey to them. In total, we sent the survey to 45 companies. %The link of the survey was sent to the email address of employees of 26 companies. We sent the link of the survey using the social media pages (LinkedIn) of employees of 10 companies because we did not have their email addresses.
We got \change{20} responses for a response rate to the survey of \change{44\%}. These 20 responses are from at least \change{17} companies.  We cannot compute the exact number because three respondents did not identify their company name.

\begin{figure*}[!ht]
\centering
\includegraphics[width=0.9\textwidth]{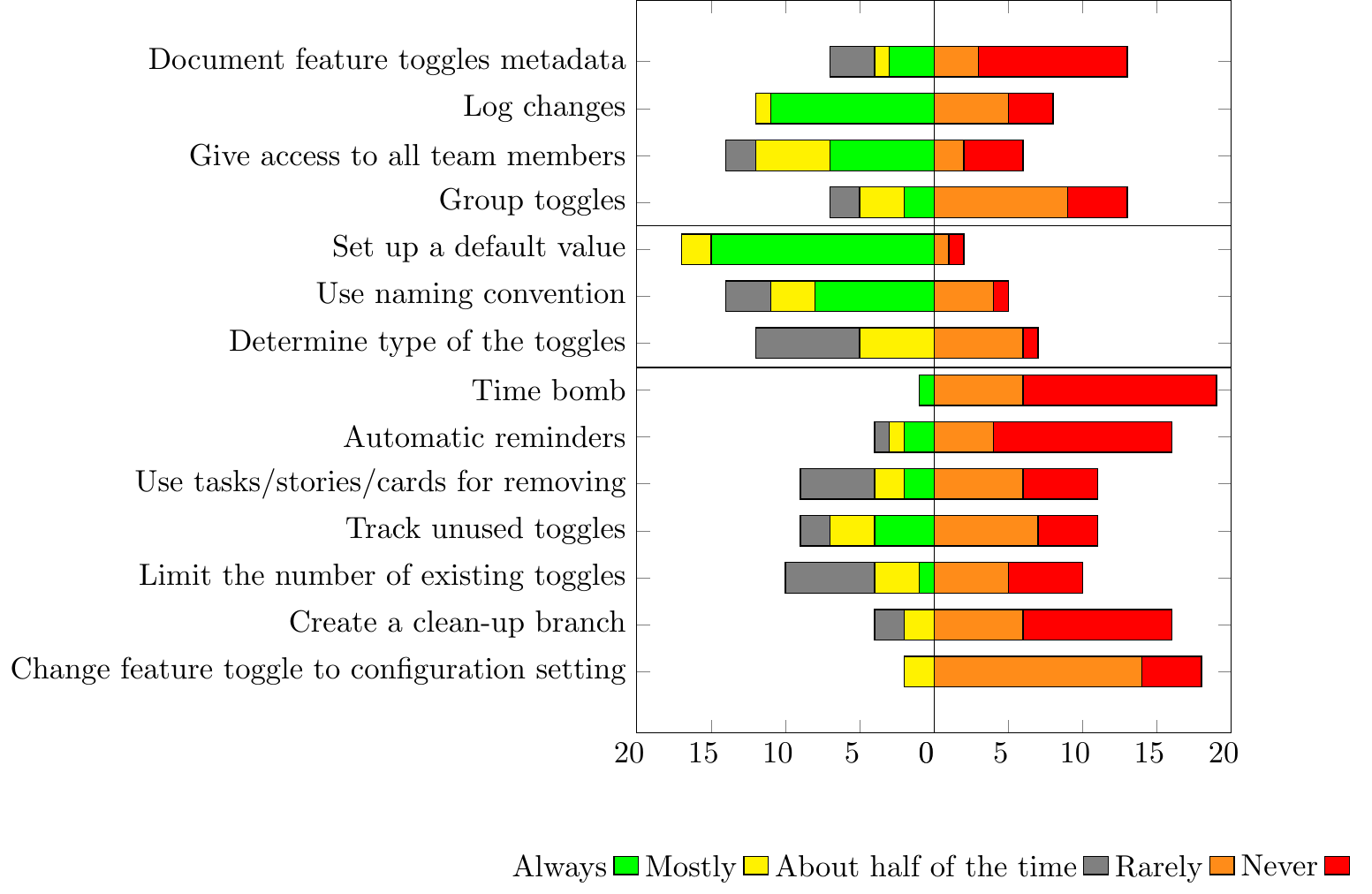}
\caption{Frequency of using the subset of feature toggle practices with Likert scale based on the survey}
\label{fig:likertchart}
\end{figure*}

As mentioned in Section \ref{methodology}, we used a Likert scale \cite{likert1932technique} with five options for 12 of the 17 practices for which a Likert scale options can be used.  In our analysis, we grouped \textit{Always}, \textit{Mostly} and \textit{About half of the time} responses and assumed the companies that selected these options use the practice. We also grouped \textit{Rarely} and \textit{Never} and assumed the company does not use the practice if the respondents selected one of these two options. The detailed result of the survey responses on the 12 questions with Likert scale options is shown in Figure~\ref{fig:likertchart}.  For the 5 remaining practices which includes all 3 implementation practices and 2 management practices, survey respondents chose from a list of provided answers or to add text to an open-ended ``other'' response.  For example, for the ``Use management systems'' (M1) respondents could chose among open or closed source systems, chose they did not use a management system or add any answer to ``other'' response. The survey questions are listed in the Appendix of the paper.

We use results from analyzing company-specific artifacts and survey responses to answer RQ2. The result of analysis of company-specific artifacts is shown in third column and the survey result is shown in the last column of Table \ref{tab1}. The frequency of usage of each practice in both company-specific artifacts and survey result is shown in this table.

Based on the survey responses, the companies which the survey respondents work have been using feature toggles for an average of 4.8 years. Among 20 respondents, 19 respondents use toggles to have gradual roll out. Nineteen respondents use toggles to support CI of partially-completed features, and 17 respondents use toggles to perform A/B testing. Fifteen respondents use toggles to have dark launches.

In the following subsection, we go through each of the categories and highlight the main findings on usage of practices. The analysis is based on the survey responses and company-specific artifacts for each category of practices. 

\subsubsection{Management Practices}
The most used practice is ``Use management systems'' (M1) based on both company-specific artifacts and survey responses. \change{However, in comparison to configuration options, Sayagh et al.  \cite{sayagh2018software} show that developers do not tend to use existing configuration frameworks.}

For the ``Determine the applicability of feature toggle'' (M4) practice, four survey respondents stated that the feature toggle is always added when a new feature is added or any feature is changed. They do not have any decision making process for using feature toggles. In companies where a feature toggle is added for each new feature, there will eventually be a large number of feature toggles so management and deletion of the toggles are more critical to prevent increased code complexity and dead code. The ``Log changes'' (M3) practice enables practitioners to follow ``Track unused toggles'' (C2) practice from clean-up category as shown in Table~\ref{relatedpractices}. If a company logs every change made on feature toggles, tracking unnecessary toggles will be easy.

\subsubsection{Initialization Practices}
 The most used practice in the initialization category based on both company-specific artifacts and survey responses is ``Set up the default values'' (I1) based on Table~\ref{tab1}. The ``Use naming conventions'' (I2) and ``Determine the type of the toggle'' (I3) are next in the rankings. The usage ranking of the practices in this category is same in both company-specific artifacts and survey responses. 
 
 ``Determine the type of the toggle'' (I3) is a practice which helps practitioners to use ``Limit the number of feature toggles'' (C3) practice in the clean-up category more efficient, as shown in Table~\ref{relatedpractices}. If the type of the toggles are pre-determined, the practitioners have a list of short-lived toggles as a suggested list of toggles to remove. Instead of checking all of the toggles, the short-lived toggles could be checked for removal.

\subsubsection{Implementation Practices}
As shown in the first column of Table \ref{tab1}, the mean of usage frequencies of implementation practices is 66\% based on company-specific artifacts and 100\% based on the survey. This category of practices is the most used practices in industry based on our result. When a company uses feature toggles, the development team implements the code of feature toggle including a mechanism to store the values of the toggle, select the type of the assigned value, and determine how to access the value. 

\change{For ``Ways of accessing the values'' (Im2), the experiences from practitioners in \cite{danpiessensfeaturetoggle}, \cite{benjamindaydevops} and \cite{robertsfeaturetoggle} as we mentioned in \textbf{Examples} of the practice in Section~\ref{practices} reflect the popularity of using objects among other ways. This can justify the survey's result, the most used way is objects (25\%). All the survey's respondents which use more than one way (45\%) use objects as one of the ways.} 

For ``Store type'' (Im3), using a configuration file is more popular than using databases in the company-specific artifacts; but the survey's responses indicate that databases and a combination of configuration files and databases are most used. We allowed respondents to add their own answer for this question, and three respondents mentioned \textit{using a third party service}, such as get values from LaunchDarkly servers. However, we did not identify this option in analyzing company-specific artifacts. We added this new store type to Table \ref{tab1} for the survey responses. \change{As we mentioned the comparisons made between using configuration files and databases in \cite{hodgsonfeaturetoggle}, \cite{meyerfeatureflags}, \cite{surfingthecode}, \cite{tiwarifeaturetoggles} and \cite{benjamindaydevops} in the \textbf{Examples} part of ``Store Type'' (Im3) practice, using databases give development team more flexibility but using configuration files is faster. These comparisons from practitioners' point of view can justify different results for the practice based on company-specific artifacts and survey. Companies may started with configuration files and after realizing the disadvantages of it, switch to using database or a combination of configuration files and database to have fast updating of values.} 

The difference between usage frequencies of practices based on the artifacts and based on the survey's responses shows that companies may change the implementation details of feature toggles over time and based on their experiences. Company-specific artifacts mentioned the feature toggles implementation details in a time of publishing the artifacts, but the survey's responses reflects the current implementation details.

\subsubsection{Clean-up Practices}

Based on the company-specific artifacts and survey's responses, the practices of the clean-up category are the least used category of practices. The mean of usage frequencies of clean-up practices are 3\% based on company-specific artifacts and 39\% based on the survey which is the lowest frequency category. 

\change{We did not find any comparison between ``track unused toggles'' (C2) and ``Change a feature toggle to a configuration setting'' (C5) with the rest of the practices in this category. Based on the comparisons done by practitioner for the rest of the practices as we mentioned in \textbf{Examples} of each practice in Section~\ref{practices}, ``Use cards/tasks/stories for removing toggles'' (C1.3) seems to be not very useful and ``Time bombs'' (C1.1) are extreme. However, ``Limit the number of feature toggles'' (C3) and ``Create a cleanup branch'' (C4) are good to follow based on the practitioners' experiences. Companies may use more than one of these practices for cleaning up the feature toggles, as we observed in survey responses.} 

\change{Among the survey responses, twelve companies are from 38 companies which we have their company-specific artifacts and their feature toggle practices are shown in Table~\ref{tab2}. We are able to compare what we found in company-specific artifacts with their responses to the survey for these twelve companies. For eight companies, there is discrepancies in survey responses and company-specific artifacts about the type of assigned values to feature toggles and ways of accessing these values in Implementation practices. There are no discrepancies found for Clean-up practices, the survey responses and observation from company-specific artifacts are aligned. Three of the companies do not have documentation for their feature toggles anymore, based on survey responses. It could be because of the management systems they have which provide documentation facilities. One of the companies followed ``Determine the applicability of feature toggles'' (M4) based on the company-specific artifacts. However, they add feature toggle for any new feature now, based on their response to the survey.}

\section{Discussion} \label{Discussion}

%In the following sub-sections, we provide analysis of the feature toggle practices usage frequency for each category \change{and compare practices when the comparison is applicable} based on company-specific artifacts and survey responses showed in Table~\ref{tab1}.

%\subsection{Management practices}

\change{Since each of the Management and Initialization category practices has positive effects on feature toggles management, we suggest practitioners to use them in their feature toggle management process, specially ``Use management systems'' (M1).  Our survey indicated that all respondents use the M1 practice because managing the added complexity and technical code of adding feature toggles is easier when the management of feature toggles is carried out in a separate system. }

\change{Each of the Implementation category practices provides practitioners more than one option to choose. We suggest practitioners select an appropriate option for type of toggles, ways of accessing those toggles and store type. In the examples under \textbf{Examples} of the practices in Section~\ref{practices}, the advantages and disadvantages of different options are provided which could guide the practitioners on selecting appropriate options.}

\change{The common goal of all practices in the Clean-up category practices is to remove feature toggles when the purpose of using the toggles is accomplished. Based on practitioners' experiences, ``Use cards/tasks/stories for removing toggles'' (C1.3) is not very useful and ``Time bombs'' (C1.1) are extreme. However, ``Limit the number of feature toggles'' (C3) and ``Create a cleanup branch'' (C4) are good to follow. We suggest, in their feature toggle development process, practitioners consider these advantages and disadvantages of clean-up practices. A practitioner can choose to use multiple of these Clean-up practices (C1-C5). They are all used to remove unused toggles with different strategies.}

%\subsection{Initialization practices}

%\subsection{Implementation practices}

%For ``Type of assigned values'', 50\% of the companies use more than one identified way (string, boolean and multivariate) based on company-specific artifacts; and based on the survey's result, 55\% of survey's respondents use more than one way. For ``Ways of accessing the values'', all of the companies that mentioned the way of accessing the values use ``managers'' based on company-specific artifacts, but 45\% of the survey's respondents use more than one of the identified ways.

%\subsection{Clean-up practices}

%Cleaning-up the toggles is one of the important activities when using feature toggles. 
%An excess of toggles is one of the problems that the development team may face\cite{rahman2016feature}. Sometimes an existing toggle that has not been used for a long time can cause severe damages. For example, the Knight Capital Group, as discussed in Section \ref{intro}, went bankrupt due to the unintended activation of an old feature toggle. Companies may neglect clean-up activities. For example, Microsoft has no centralized process to clean-up the feature toggles; each team takes care of their feature toggles and decide when to remove them \cite{microsoftexperiment}. As discussed earlier, a negative consequence of using feature toggles is increased complexity and dead code, so removing unneeded toggles is necessary.

\section{Limitations} \label{limitation}

In this section, the limitations of research are discussed.
\subsection{Finding artifacts}

In Step One, we used a keyword search based upon five keywords to find grey literature and peer-reviewed papers and selected artifacts that were related to the use of feature toggles in software development. We also followed links and references to other artifacts in selected artifacts. We may have missed artifacts. 

In Step Three, we searched for company-specific artifacts based the companies found in the initial artifacts. Data from companies who have not shared their results on the Internet are not included in our study.  

\subsection{Identification of Practices and Categories}

In Step Two, we did not use any automatic technique or tools to identify practices. We may have missed some practices which were mentioned implicitly in artifacts.

Another limitation is the lack of specific examples by companies of using feature toggle practices. Practitioners mentioned most of the practices with no concrete example.

In addition, testing practices are not identified and mentioned in the list of practices. Testing of the system which has feature toggles has different aspects, such as unit testing of feature toggles, testing all combination of feature toggles enabling and disabling, and testing dependent feature toggles. Another study should be conducted to cover testing concerns and practices when a development team use feature toggles.

\subsection{Extraction of Practice Usage from Company-specific Artifacts}

In Step Four, we reviewed company-specific artifacts to extract feature toggle practices usage. If the practice was not mentioned in the artifacts, we cannot conclude that the company does not use the practice. To overcome this limitation, we conducted the survey to gather more information about usage of feature toggle practices in companies.

\subsection{Conducting Survey}

In Step Five, we found contact information of individuals associated with company-specific artifacts or who were release managers or developers of the companies. The contact information for some of the individuals could not be found or was old and out of date. To overcome this limitation, we found contact information of current development team members, such as release manager or developers of the companies, using company website or social media pages, such as LinkedIn. Additionally, the small sample size of the survey was a limitation. 

\section{Conclusion and Future Work} \label{conclusion}

Feature toggles are a technique often used by companies who practice CI/CD to integrate partially-completed features into the code, conduct a gradual roll out, and/or to perform experiments. However, the development practices used by these organizations have not been enumerated in prior research. We performed qualitative analysis of 99 artifacts from grey literature and 10 peer-reviewed papers. We identified 17 feature toggle practices in four categories: Management practices (6), Initialization practices (3), Implementation practices (3), and Clean-up practices (5). We also quantified the frequency of usage of these identified practices in industry by analyzing company-specific artifacts and conducting a survey. 

The most popular practice in each category is consistent across the company-specific artifacts and survey responses. We observed that all of the survey's respondents ``Use a management system'' (M1) to create and manage feature toggles in their code. ``Document feature toggle's metadata'' (M2), ``Log changes'' (M3), and ``Set up the default values'' (I1) are three additional highly-used practices in industry based on company-specific artifacts. The least used category of practices is Clean-up practices, even though cleaning-up the feature toggles helps with managing the added complexity to the code and removing dead code. Inattention to removing feature toggles can cause severe problems, such as what happened to Knight Capital Group. 

The feature toggle development practices discovered and enumerated in this work could raise practitioners' awareness of feature toggle practices and their usage in industry. Using the result of this paper could help practitioners to better use feature toggles in their projects, which was the goal of doing this work. We got feedback from some of the respondents to the survey that the clean-up practices seem interesting, and they decided to use identified practices in their companies after participating in the survey. 

  The identified feature toggle practices discovered through this work can enable future quantitative analysis to automatically identify practice use in code repositories. Also a future study can be conducted on the empirical analysis of the effectiveness of the identified practices in repositories which practices are use. %The  identified  feature  toggle  practices  in  this paper  could  be  used  in  future  quantitative  analysis  researches about  feature  toggles  since  the  result  of  qualitative  analysis  researches could be the start point to conduct following quantitative analysis researches  in  the  same  domain \cite{menzies2019assessing}. 
  %In addition, a subset of these identified practices can be identified automatically by analysis of the code repository of the projects, so automatic identification of these practices is one of the future works. 
  Additional future work involves the automatic identification of feature toggle bad smells in the code, such as unused feature toggles, nested feature toggles, and development of a tool to automatically refactor the code when bad smells are identified.
  Also, the quality of parts of the code which is activate or deactivate by feature toggles is one of the concerns mentioned by practitioners \cite{sowafeaturebits}. Studying the impact of using feature toggles on code quality, such as high cohesion and low coupling, could also be a future work.
  
  %\change{In addition, because of identifying practices that have overlap with the practices related to configuration options \cite{sayagh2018software}, and the suggestion of Meinicke et al.  \cite{meinicke2019exploring} to transfer solutions from configuration option community to feature toggle community, another future work could be check the applicability of reported solutions for configuration option engineering on feature toggle management. }

\bibliographystyle{unsrt}
\bibliography{bibliography.bib}

\newpage
\section*{Appendix}
The survey questions are as follows:
\begin{enumerate}[labelindent=0pt]
    \item What is your company name?
    \item How long has your team used feature toggles? %( Estimation is fine if you do not know the exact time)
    \item What feature toggle management system is used by your team? (Check all that apply). Options: Closed source custom system maintained by the company; Open source custom system maintained by the company; Third party (e.g. LaunchDarkly), Open source but not maintained by the company; None; Other.
    \item For what purpose(s) does your team use feature toggles? (Check all that apply). Options: Support CI of partially-completed features; Dark launches; A/B testing; Gradual rollout; Other.
    \item Does your team make decision about using feature toggles for each feature? Options: Yes. The team checks to find if using a feature toggle is necessary for the new feature; No. The feature toggle is always added when a new feature is added; Other.
    \item How often does your team do the following management practices? Options: Always, Mostly, About half of the time, Rarely, and Never.
        \begin{itemize}
            \item Document feature toggle's metadata (spreadsheet, etc) to manage data about feature toggles. (i.e. the owner of the toggle, the current value (on, off), the current status (to remove, keep) and the time of its creation).
            \item Logging changes to toggle values/configurations (e.g. who changes which toggle and when, etc.).
            \item Grouping toggles together in any way to simplify management or giving permissions (i.e. related toggles, other).
            \item Allowing all team members (i.e. Q\&A team) to have access to feature toggles and can make changes.
        \end{itemize}
    \item How often does your team do the following initialization practices? Options: Always, Mostly, About half of the time,	Rarely and Never.
        \begin{itemize}
            \item Determining the type (permission toggle, ops toggle, release toggle, experiment toggle, short-lived toggle, long-lived toggle) of the toggle at design step. (More information about types of toggles: \url{https://goo.gl/4okG5Y})
            \item Using naming conventions for toggles (similar to variable and function naming conventions).
            \item Setting up a default value for toggle if toggle value is not found (i.e. toggle is off if its value is not found in the code).
        \end{itemize}
    \item How are the values of the toggles stored? (Check all that apply) Options: Configuration files; Databases; Other.
    \item How are the values are assigned to the toggles in the system? (Check all that apply) Options: Assigned boolean values (True, False); Assigned multivariate values (e.g. Red, Yellow, Blue); Assigned string values (e.g. "disable-flash-3d", "enabled-flash-3d"); Other.
    \item How does a developer access the toggle value in the code? (Check all that apply) Options: Value is accessed by checking a primitive data type (e.g. enableMyFeature == true),; Value is accessed through an object representing a toggle (e.g. MyFeature.isActive()); Value is accessed through a toggle manager/mapping from key to value (e.g. Dictionary); Other.
    \item How often does your team do the following clean-up practices? Options: Always, Mostly, About half of the time,	Rarely and Never.
        \begin{itemize}
            \item Limiting the number of existing toggles in the code.
            \item Build or test failing if a toggle is not deleted by a specified date (Time bomb).
            \item Automatic reminders near date to delete the toggle.
            \item Using tasks/stories/cards for removing toggles.
            \item Creating a clean-up branch for removing toggle points at the time of creation of the toggle.
            \item Tracking unused toggles for removal.
            \item Changing feature toggle to configuration setting to keep it in the code.
        \end{itemize}
\end{enumerate}

\begin{table*}[h] %btp to have at the middle of text
\setlength{\tabcolsep}{3pt}
\caption{38 Companies and their usage of identified practices from company-specific artifacts}
\begin{tabular}{@{} r *{18}{|c}}
\multicolumn{1}{c|}{ } & \multicolumn{6}{c|}{Management} &
\multicolumn{3}{c|}{Initialization} &
\multicolumn{3}{c|}{Implementation} &
\multicolumn{5}{c}{Clean-up} \\ \hline
  Company &
  \rotatebox[origin=c]{90}{Use management systems} &
  \rotatebox[origin=c]{90}{Document feature toggle's metadata} &
  \rotatebox[origin=c]{90}{Log changes} &
  \rotatebox[origin=c]{90}{Give access to team members} &
  \rotatebox[origin=c]{90}{Determine  applicability  of feature toggle} &
  \rotatebox[origin=c]{90}{Group the feature toggles} &
  \rotatebox[origin=c]{90}{Set up the default values} &
  \rotatebox[origin=c]{90}{Use naming convention} &
  \rotatebox[origin=c]{90}{Determine  the  type  of  the toggle} &
  \rotatebox[origin=c]{90}{Type   of   assigned   values} &
  \rotatebox[origin=c]{90}{Ways  of  accessing  the  values} &
  \rotatebox[origin=c]{90}{Store  type} &
  \rotatebox[origin=c]{90}{Track unused toggles} &
  \rotatebox[origin=c]{90}{Add expiration date} &
  \rotatebox[origin=c]{90}{Change a feature toggle to a configuration setting} &
  \rotatebox[origin=c]{90}{Limit  the  number  of  feature toggles} &
  \rotatebox[origin=c]{90}{Create a cleanup branch} \\
  \hline
  Airbnb & \checkmark & \checkmark & \checkmark & & & & \checkmark & & & & \checkmark & & & & & &\\
  \hline
Apiary & \checkmark & \checkmark & \checkmark & \checkmark & & & \checkmark & & & \checkmark & \checkmark & & & & & & \\
\hline
AppDirect & \checkmark & \checkmark & \checkmark & & & & \checkmark & & & \checkmark & \checkmark & & & & & & \\
\hline
Behalf & \checkmark & \checkmark & \checkmark & & & & \checkmark & & & \checkmark & \checkmark & & & & & & \\
\hline
CircleCI & \checkmark & \checkmark & \checkmark & \checkmark & & & \checkmark & & & \checkmark & \checkmark & & & & & & \\
\hline
Checkr & \checkmark & & \checkmark & & & & & & & \checkmark & \checkmark & \checkmark & & & & & \\
\hline
commercetools & \checkmark & \checkmark & \checkmark & & \checkmark & & & & & \checkmark & & & & \checkmark & & & \\
\hline
Domain & \checkmark & \checkmark & & & & & \checkmark & & & \checkmark & & & & & & & \\
\hline
DropBox & \checkmark & & \checkmark & & & & & & & \checkmark & \checkmark & \checkmark & \checkmark & & & & \\
\hline
Envoy & \checkmark & \checkmark & \checkmark & \checkmark & & & \checkmark & & & \checkmark & \checkmark & & & & & & \\
\hline
Etsy & \checkmark & & & & & \checkmark & \checkmark & & & \checkmark & \checkmark & \checkmark & & & & & \\
\hline
Facebook & \checkmark & \checkmark & & & & & & & & \checkmark & \checkmark & \checkmark & & & & & \\
\hline
FINN.no & \checkmark & \checkmark & \checkmark & & \checkmark & & \checkmark & & & \checkmark & \checkmark & & & & & & \\
\hline
Flickr & & \checkmark & & & & & & & & \checkmark & \checkmark & \checkmark & & & & & \\
\hline
GoPro & \checkmark & \checkmark & \checkmark & & & \checkmark & \checkmark & & & \checkmark & \checkmark & & & & & & \\
\hline
Google Chrome & & \checkmark & & & & & \checkmark & & & \checkmark & \checkmark & \checkmark & & & & & \\
\hline
IBM & & & & & & & & \checkmark & & \checkmark & & \checkmark & & & & & \\
\hline
Instagram & \checkmark & \checkmark & \checkmark & \checkmark & & & & & & \checkmark & \checkmark & & & & & &\\
\hline
InVision & \checkmark & \checkmark & \checkmark & & \checkmark & & \checkmark & \checkmark & \checkmark & \checkmark & \checkmark & & & \checkmark & & & \\
\hline
Librato & \checkmark & & & & & & & & & \checkmark & \checkmark & \checkmark & & & & & \\
\hline
Lyris & & & & & \checkmark & & \checkmark & \checkmark & & & & \checkmark & & \checkmark & & & \\
\hline
Main Street Hub & \checkmark & \checkmark & \checkmark & & & & \checkmark & & & \checkmark & \checkmark & & & & & & \\
\hline
Microsoft & \checkmark & \checkmark & & & \checkmark & & \checkmark & & & \checkmark & \checkmark & \checkmark & & & & & \\
\hline
Outbrain & & & & & \checkmark & & & & & & & \checkmark & & & & & \\
\hline
Pinterest & \checkmark & \checkmark & \checkmark & & & & \checkmark & & & & \checkmark & \checkmark & & & & & \\
\hline
Rally Software & & \checkmark & & & & & & & & & & & & \checkmark & & & \\
\hline
Reddit & \checkmark & & & & & & \checkmark & \checkmark & & \checkmark & \checkmark & \checkmark & & & & & \\
\hline
Slack & \checkmark & & & & & & & & & & & & & \checkmark & & & \\
\hline
Soluto & \checkmark & & & & & & & \checkmark & & \checkmark & & & & & & & \\
\hline
Surfline & \checkmark & \checkmark & \checkmark & \checkmark & & & \checkmark & & & \checkmark & \checkmark & & & & & & \\
\hline
ThoughtWorks  & \checkmark & & & \checkmark & & & & & & \checkmark & \checkmark & \checkmark & & \checkmark & & & \\
\hline
thredUP & \checkmark & \checkmark & \checkmark & & \checkmark & & \checkmark & & & \checkmark & \checkmark & & & & & & \\
\hline
Travis-CI & \checkmark & & & & & & & & & \checkmark & \checkmark & & & & & & \\
\hline
Twilio & \checkmark & \checkmark & \checkmark & & & & \checkmark & & & \checkmark & \checkmark & & & & & & \\
\hline
Upserve & \checkmark & \checkmark & \checkmark & & \checkmark & & \checkmark & & & \checkmark & \checkmark & & & & & & \\
\hline
Visma & \checkmark & & & & & & & & & \checkmark & & & & & & & \\
\hline
WePay & \checkmark & \checkmark & \checkmark & & & & \checkmark & & & \checkmark & \checkmark & & & & & & \\
\hline
Wix & \checkmark & \checkmark & \checkmark & \checkmark & & & & & & \checkmark & & \checkmark & & & & & \\
  \hline
Total (38) & 32 & 25 & 21 & 7 & 8 & 2 & 22 & 5 & 1 & 32 & 28 & 15 & 2 & 6 & 0 & 0 & 0 \\
  \hline
\end{tabular}
\label{tab2}
\end{table*}

\end{document}